\documentclass[11pt]{article}
\usepackage[T1]{fontenc}		
\usepackage[ansinew]{inputenc}
\usepackage{amsmath}
\usepackage{amssymb}
\usepackage{slashed}

\usepackage{graphicx}
\usepackage{subfigure}
\usepackage[usenames]{color}
\usepackage{hyperref}

\newcommand{\be}{\begin{equation}}
\newcommand{\ee}{\end{equation}}
\newcommand{\beqa}{\begin{eqnarray}}
\newcommand{\eeqa}{\end{eqnarray}}
\newcommand{\bseq}{\begin{subequations}}
\newcommand{\eseq}{\end{subequations}}

\newcommand{\pd}{\partial}
\newcommand\m{\mu}
\renewcommand\a{\alpha}

\newcommand\n{\nu}
\renewcommand\l{\lambda}
\renewcommand\b{\beta}	
\renewcommand\r{\rho}
\newcommand\di{\mathrm{d}}
\newcommand\s{\sigma}

\title{\sc{\Huge Gravitational Radiation in 
Ho\v rava Gravity}}

\author{Diego Blas,\!$^a$ Hillary Sanctuary,\!$^{b}$\vspace{.2cm}\\
\normalsize\llap{$^a$}
 \it FSB/ITP/LPPC,
 \'Ecole Polytechnique F\'ed\'erale de Lausanne,\\
 \normalsize\it CH-1015, Lausanne, Switzerland\\
\normalsize\llap{$^b$} \it  D\'epartement de Physique Th\'eorique, Universit\'e de Gen\`eve, \\
\normalsize\it CH-1211 Geneva, Switzerland.}

\begin{document}

%%%%%%%%%%%%%%%%%%%%%%%%%
\maketitle

\begin{abstract}
We study the radiation of gravitational waves by self-gravitating 
binary systems in the low-energy limit of Ho\v rava gravity.
We find that the predictions for the energy-loss formula of General Relativity are modified already for Newtonian sources:
the quadrupole contribution
is altered, in part due to the different speed of propagation of the tensor modes; furthermore, there is a monopole contribution
 stemming from an extra scalar degree of freedom. A dipole contribution only appears at higher post-Newtonian order. 
  We use these findings to constrain the low-energy action of  Ho\v rava gravity by
comparing them with the radiation damping observed for binary pulsars. Even if 
this comparison is not completely appropriate --  since compact objects cannot be described
within the post-Newtonian approximation -- it represents an order of magnitude estimate.
In the limit where the post-Newtonian metric coincides with that of General Relativity, our
energy-loss formula provides the strongest constraints for  Ho\v rava gravity at low-energies.
\end{abstract}
\newpage
\tableofcontents
%%%%%%%%%%%%%%%%%%%%%%%%%%%%%%%

%%%%%%%%%%%%%%%%
\section{Introduction}
%%%%%%%%%%%%%%%%

General Relativity (GR) continues to stubbornly agree with every observation related to gravity \cite{Will:2005va}. 
This would be extremely desirable if the theory could be merged with quantum mechanics in a straightforward way. Unfortunately, 
 the current situation is far from this: 
the search for a consistent theory of quantum gravity
 remains elusive and there is no experimental guidance to shed light on it. Furthermore,  the cosmological constant problem
  aside\footnote{One may argue that the cosmological constant problem is a hint towards the actual theory of quantum gravity, and that a successful framework of quantum gravity should provide a mechanism to explain this phenomenological observation. We do not address this particular issue here.}, the success
of GR as a low-energy effective field theory (EFT)  points towards the Planck mass $M_P\approx 10^{19}\
\mathrm{GeV}$ as
the physical frontier where one expects to learn anything about quantum gravity. If the preceding arguments are 
realized in Nature, experimental information about quantum gravity will indeed be sparse in the foreseeable future.

More interesting for phenomenology are the proposals for ultraviolet (UV) completions of GR where the previous logic fails.
 These include models of gravitation with a low-energy cutoff beyond which GR ceases to be valid \cite{ArkaniHamed:1998rs,Dvali:2007hz}. 
 If this cutoff scale is as low as the $\mathrm{TeV}$, these proposals  may have interesting phenomenology and 
 may even be relevant for the resolution of the hierarchy problem.
Another recent proposal in this category is Ho\v rava gravity  \cite{Horava:2009uw,Blas:2010hb}. This framework yields a concrete UV completion of GR, with effects that may permeate basically any 
gravitational experiment. It is on the implications of Ho\v rava gravity for gravitational radiation that we pursue in this paper.

Essentially, Ho\v rava's proposal consists of considering 
the existence of a preferred time-foliation of spacetime. Assuming the presence of this absolute structure, the GR
Lagrangian can be supplemented
by operators which render it power-counting renormalizable without destroying the unitarity of the theory \cite{Horava:2009uw}.
The result is a non-relativistic theory of 
quantum gravity  \cite{Blas:2010hb}  (in the sense that it is Lorentz-violating).
The preferred foliation is in principle detectable at any energy scale, and it is not surprising that this approach (which is designed
to cure the unsatisfactory behaviour of GR at distances of the order $M_P^{-1}\approx 10^{-33}\ cm$)  generically also modifies the theory
at large distances\footnote{A counterexample to this argument can be found in \cite{Horava:2010zj}. However, it is
not clear how GR is recovered at large distances in this proposal.}  \cite{Blas:2010hb}.
Among the different implementations of Ho\v rava's idea, 
 we consider the so-called ``healthy extension''  \cite{Blas:2009qj}. This version possesses a stable
Minkowski background where the issues about strong coupling appearing in other approaches are absent. 
Furthermore, variants of Ho\v rava's original proposal can be retrieved for a particular limit of this (generic and stable) case \cite{Blas:2009qj}.

The low-energy (large-distance) sector of the theory is encoded into a scalar field $\varphi$,
 called\footnote{From Greek $\chi\rho o\nu o \varsigma$ -- time. The khronon is also known as the ``T-field''  \cite{Jacobson:2010mx}.} the ``khronon'', that characterizes
the foliation structure and interacts with a metric field. We refer to this 
low-energy scalar-tensor theory as ``khronometric'' theory \cite{Blas:2010hb}. 
The extra scalar field $\varphi$ turns out to be massless, and its presence modifies most
of the predictions of GR, including the parametrized-post-Newtonian (PPN) parameters  \cite{Blas:2010hb,Blas:2009qj,Blas:2009ck} and cosmological phenomena \cite{ArmendarizPicon:2010rs,Kobayashi:2010eh}.  These modifications differ from those of standard scalar-tensor
theories \cite{Will:2005va,Will:Book}. They are close, however, 
to the predictions of Einstein-aether theory (or \ae-theory for short) \cite{Jacobson:2008aj}. This is not surprising since both theories incorporate a field whose 
expectation value violates Lorentz invariance (a unit timelike  vector in the case of Einstein-aether, and $\varphi$ in our case),
 and are otherwise generic. 
It can be shown that the khronon $\varphi$ corresponds to the hypersurface-orthogonal mode of \ae-theory,  and many 
of the predictions of both theories are indeed identical \cite{Blas:2010hb,Jacobson:2010mx}.
The PPN parameters derived from \ae-theory and khronometric theory
 restrict the parameter space of those theories but are otherwise in agreement with current observations.
 Thus, both (low-energy) models represent interesting 
alternatives to GR, which, furthermore, have a high energy cutoff . The further advantage of khronometric theory is that beyond this
energy cutoff
 there is a known UV completion, in the form of Ho\v rava gravity.

The aim of this paper is to further constrain khronometric theory based on the 
 loss of energy due to the emission of gravitational waves (GWs) from a binary self-gravitating system.  This is 
 a relevant test for gravitational theories given its sensitivity to the way gravity propagates (e.g. the degrees of freedom and corresponding properties), and also to
 the strong-field regime since known astrophysical sources of GWs tend to have strong gravitational self-energies \cite{Will:Book,Blanchet:2006zz,Maggiore:1900zz}.
 The confirmation of GR's famous quadrupole formula in the damping of a binary pulsar's orbit is indeed one of its triumphs 
 \cite{Taylor:1982zz,Weisberg:2004hi}. Radiation tests have also been used in the past 
 to constrain possible modifications of GR \cite{Will:2005va,Damour:1992we,Cannella:2009he}.
 A priori for both \ae-theory and khronometric theory, one expects this radiation formula to be modified due to
 a different speed of propagation of 
 the tensor modes  and the presence of new propagating fields. These modifications imply new ways to constrain the parameter space of the theory, independently of PPN and cosmological considerations.
While the above expectations have been verified for \ae-theory in the weak-field regime in \cite{Foster:2006az}, the constraints obtained are not final since the astrophysical systems for which radiation damping has been observed are not in the weak-field regime \cite{Will:2005va}.  The incorporation of
 strong-field effects in the Einstein-aether began in \cite{Foster:2007gr}.

We focus on the radiation formula in the post-Newtonian (weak field, 
slow-motion and weakly stressed \cite{Blanchet:2006zz}) regime
of khronometric theory. This restriction is interesting for two reasons. 
First, we find deviations from GR's quadrupole formula already at leading
order.
(This is similar to what happens in \ae-theory, as computed in \cite{Foster:2006az}.)
Second, and from a purely pragmatic point of view, many of the formulae we present in this paper are useful for the 
phenomenologically relevant situation of compact sources.
 First results relevant for the study of gravitational radiation from these systems 
 include the black hole solutions derived in \cite{Barausse:2011pu,BlasSibi},
and those for neutron stars in \ae-theory \cite{Eling:2007xh}. The use
of binary pulsar observations to constraint Ho\v rava gravity
was  suggested in \cite{Kimpton:2010xi}.

To extract information about the damping of the orbit of a binary self-gravitating system from the emission of GWs, we 
take advantage of the fact that khronometric theory is semi-conservative (in the language of \cite{Will:Book}). Then, for the bound system there exists a conserved energy that decreases due to the emission of gravitational radiation. By computing the energy flux at infinity, we can derive the flux of energy lost  by the binary. 
Under the assumption that this energy is extracted entirely  from the orbital motion of the binary,
 the subsequent damping of the orbits can be computed using Kepler's third law.  This assumption has been 
 tested to lowest order in GR \cite{Maggiore:1900zz}, and is plausible for khronometric theory. 

This work is structured as follows.
In  Sec.~\ref{sec:lowEaction}, we define the action for khronometric theory and the equations of motion relevant for low-energy phenomenology.  
Sec.~\ref{sec:perturbation} is devoted to the linearized equations for the fields far away from the source (far-zone). 
In Sec.~\ref{sec:source},  we study the  conserved properties of the source relevant for the post-Newtonian (PN) calculation.
We derive the explicit expressions for the the different waveforms, up to and including the first PN order  corrections in Sec.~\ref{sec:waveforms}.
In Sec.~\ref{sec:energy}, we  determine the formula for the average power loss in
GWs. This formula is evaluated for a Newtonian system of two point-masses in Sec.~\ref{sec:binary}, where the Peters-Mathews parameters
for khronometric theory are derived. We summarize our results and conclude in Sec.~\ref{sec:discussion}. 
 Appendix \ref{app:PPN} contains a derivation of the PPN parameters for our model (whose full expressions appear here for the first time).  
 Appendix \ref{app:monopole} compares the monopole contribution, or lack thereof, in both khronometric theory and \ae-theory for a particular choice of parameters.
 Finally, Appendix \ref{ap:energy} provides a summary of the notion of energy relevant for our study.

 \subsection*{Conventions} 
 
We use the $(+---)$ signature.  For an arbitrary expression $X$, the overbar $\bar{X}$  denotes the part of $X$ linear in perturbations. 
 The superscript $X^{NL}$ is the non-linear part of $X$, i.e. $X^{NL}\equiv X-\bar X$.  The dot  $\dot{X}$ denotes the derivative of $X$ with respect to time.  Greek indices refer to spacetime, whereas Latin indices refer to space only. Repeated Latin indices are to be summed, e.g. $X_{ii}\equiv \delta^{ij}X_{ij}$. We define the symmetrization of indexes as $T_{(ij)}\equiv \frac{1}{2}\left(T_{ij}+T_{ji}\right)$. We choose units where $c=\hbar=1$.

%%%%%%%%%%%%%%%%
\section{Action for khronometric theory at low energies} \label{sec:lowEaction}
%%%%%%%%%%%%%%%%

As outlined in the introduction, Ho\v rava gravity is based on the existence of an absolute time foliation of spacetime.
This allows for the GR Lagrangian to be supplemented with higher dimensional operators that render the theory power-counting
renormalizable  \cite{Horava:2009uw}. 
 These operators are suppressed by a scale $M_*$ whose 
 magnitude is constrained by various phenomenological tests. The most stringent of these tests comes
 from absence of deviations from Newton's law at short distances \cite{Blas:2010hb} which implies that
  $M_*\gtrsim (10\ \m m)^{-1}\sim 10^{14}  \ \mathrm{Hz}$ \cite{Will:2005va,Blas:2010hb}. Thus,  these
  higher dimensional operators are irrelevant for the binary systems of interest\footnote{As an example, the famous
PSR 1913+16 binary pulsar has a characteristic frequency of $10^2 \ \mathrm{Hz}$ \cite{Will:2005va}. We assume that the speed of propagation of all the modes is similar to the speed of light.  We comment on this assumption when we derive the energy-loss  formula in Sec. \ref{sec:energy}.} and we neglect them in the
following.
The presence of a preferred foliation also has consequences at energy scales below $M_*$. Indeed,  
at low-energies new operators appear (compared to GR) that are compatible
with the group of gauge invariance preserving the preferred foliation, i.e. the foliation-preserving diffeomorphism 
\cite{Horava:2009uw,Blas:2010hb}.  Renormalization group arguments imply that these relevant operators 
should be added to the GR action, which has been done in the  St\"uckelberg (or covariant) formulation of the theory in \cite{Blas:2010hb,Blas:2009yd}. 
In this formulation, the preferred-time foliation corresponds to the expectation value of a scalar field $\varphi$ called the ``khronon''.  This field is such that the normal to the surfaces of constant field is timelike,
\be
\label{eq:timel}
{\pd_\m \varphi\, \pd^\m\varphi}>0.
\ee
The action of the theory 
is invariant under diffeomorphisms, and Lorentz invariance is broken by condition (\ref{eq:timel})  in a spontaneous way.   Also, the
action must be endowed 
with invariance under
field reparametrizations 
\be
\label{eq:repr}
\varphi\mapsto f(\varphi),
\ee 
which follows from our requirement of a preferred time-foliation as opposed to a preferred time.
It corresponds to the time reparametrization invariance of the theory in the original formulation of   \cite{Horava:2009uw}.
   The invariance under the transformations (\ref{eq:repr}) is readily implemented by making
 the action depend on $\varphi$ through the combination
\be
\label{eq:u}
u_\m\equiv \frac{\pd_\m \varphi}{\sqrt{\pd_\r \varphi \pd^\r\varphi}}.
\ee 
Clearly, $u_\m$ is non-singular whenever condition (\ref{eq:timel}) is satisfied.
 Notice also that $u_\m$ is a unit timelike vector.
 
The low-energy action for the healthy extension of Ho\v rava gravity corresponds to the most
general action describing the coupling of $\varphi$ with a metric field $g_{\m\n}$ at low-energies 
and compatible with the aforementioned invariances \cite{Blas:2010hb}.  It is given by
\be
\begin{split}
\label{covar22}
S=&
-\frac{{M_b}^2}{2}\int\di^4 x\sqrt{-g}\Big[R
+K^{\m\n}_{\phantom{\m\n}\s\r}\nabla_\m u^\s\nabla_\nu u^\rho\Big]\;+S_m,
\end{split}
\ee
where $M_b$ is an arbitrary mass parameter to be related to the Planck mass,
$$
K^{\m\n}_{\phantom{\m\n}\s\r}=\beta\, \delta^\m_\r \delta^\n_\s+\lambda\, \delta^\m_\s \delta^\n_\r+\alpha\, u^\m u^\n g_{\s\r},
$$
and $\a$, $\b$ and $\lambda$ are free dimensionless constants\footnote{Note that the parameter $\l$ corresponds to $\l'$
 in the notations of \cite{Blas:2010hb}. It also differs from the $\l$ defined in \cite{Horava:2009uw}.}.
We also introduce a term  $S_m$ in Eq.~(\ref{covar22})  representing the matter component of the theory. 
We assume that matter is universally coupled to the metric $g_{\m\n}$, which enforces
 the weak equivalence principle \cite{Will:2005va}. This action defines what we call ``khronometric theory''. 
For later convenience, we introduce
$$
S_{\chi}\equiv -\frac{{M_b}^2}{2}\int\di^4 x\sqrt{-g}K^{\m\n}_{\phantom{\m\n}\s\r}\nabla_\nu u^\rho\nabla_\m u^\s=
 -\frac{{M_b}^2}{2}\int\di^4 x\sqrt{-g}K^{\m}_{\phantom{\m}\s}\nabla_\mu u^\s,
$$
where 
$$
K^{\m}_{\phantom{\m}\s}\equiv K^{\m\n}_{\phantom{\m\n}\s\r}\nabla_\n u^\r= K^{\n\m}_{\phantom{\m\n}\r\s}\nabla_\n u^\r. 
$$
is used to compactify notation. 

Khronometric theory can be considered on its own as an alternative to GR with an extra scalar field, independently of quantum gravity motivations.   This
approach is similar to the way Einstein-\ae ther theories
 are constructed.  The only difference is that the vector $u_\m$ is taken to be a generic timelike vector in \ae-theory \cite{Jacobson:2008aj}, meaning that it has more degrees of freedom than in the khronometric case.  It also implies an extra term in 
the generic action with respect to Eq. (\ref{covar22}).  This extra term can be absorbed by the ones in
 action (\ref{covar22}) for hypersurface orthogonal vectors, i.e. whenever $u_\m$ satisfies Eq. (\ref{eq:u}). 
 Khronometric theory and \ae-theory share the nice feature of
having a high energy cutoff.
The advantage of the former is that  a UV completion in the form of Ho\v rava gravity is known.

Let the khronon and matter energy-momentum tensors be, respectively,
$$
T^{\chi}_{\m\n}\equiv\frac{2}{\sqrt{-g}}\frac{\delta S_{\chi}}{\delta g^{\m\n}}, \quad T_{\m\n}^m
\equiv\frac{2}{\sqrt{-g}}\frac{\delta S_m}{\delta g^{\m\n}}.
$$
The explicit expression for $T_{\m\n}^{\chi}$ reads
\be
\begin{split}
M_b^{-2}T^{\chi}_{\m\n}&=-\nabla_\r\left(K_{(\m\n)}u^\r+K^\r_{\phantom{\r}(\m}u_{\n)}-
K^{\phantom{(\m}\r}_{(\m}u_{\n)}\right)+\frac{1}{2} g_{\m\n} K^\r_{\phantom{\m}\s} \nabla_\r u^\s\\
&+\alpha \,a_\m a_\n+2 \nabla_\r K^\r_{\phantom{\r}(\m}u_{\n)}-u_\m u_\n
u^\s \nabla_\r K^\r_{\phantom{\r}\s}-2\alpha\, a_\s u_{(\m}\nabla_{\n)}u^\s+\alpha \,a^\r a_\r
u_\m u_\n,\nonumber 
\end{split}
\ee
where we have introduced the notation
$$a_\m\equiv u^\r \nabla_\r u_\m.$$
 The equations of motion derived from varying the action with respect to the metric are 
\be
\label{eq:eom}
{\mathcal Q}_{\m\n}\equiv G_{\m\n}-M_b^{-2}\left(T^{\chi}_{\m\n}+T_{\m\n}^m\right)=0.
\ee
The equation of motion for the khronon field is
\be
\label{eq:eomk}
\mathcal Q_\chi\equiv\nabla_\m J^\m\equiv \nabla_\m\left(\frac{1}{\sqrt{X}}\mathcal P^{\m\n}\left[\nabla_\sigma K^\sigma_{\phantom{\sigma}\n}-\alpha\, a_\sigma \nabla_\n u^\sigma\right]\right)=0,
\ee
where 
\be
\mathcal P^{\m\n}\equiv\left(g^{\m\n}-u^\m u^\n\right).
\ee
As usual, this equation follows from the covariant conservation of the khronon 
energy-momentum tensor.  That it can be represented as the conservation of a current is a consequence of the invariance of the theory under reparametrizations of the khronon given by Eq.~(\ref{eq:repr})  \cite{Blas:2009yd}.

%%%%%%%%%%%%%%%%%%%%%%%%%%%%%%%%%%%
\section{Equations of motion in the far-zone} \label{sec:perturbation}
%%%%%%%%%%%%%%%%%%%%%%%%%%%%%%%%%%%

The physical system of interest for radiation damping consists of an isolated self-gravitating astrophysical source.
By this we mean that there is a region of spacetime surrounding the source where the fields
acquire their background values plus small perturbations.  Thus,
there exists a coordinate frame where the metric in this region satisfies\footnote{In this section, Greek indices are manipulated with the Minkowski metric.}, 
\be
g_{\m\n}=\eta_{\m\n}+h_{\m\n},\nonumber
\ee
with $\vert h_{\m \n} \vert \ll 1$.  For the khronon field, we fix the parametrization invariance (\ref{eq:repr})
 by  working with a time coordinate corresponding to the
background of the field. Thus, we expand it as
\be
\varphi= t+ \chi,\nonumber
\ee 
where $\vert \chi \vert \ll t$.  It is easy to verify that the background fields are indeed solutions of the equations of motion (\ref{eq:eom}) in the absence of matter. To derive the flux of energy lost by this system, it is enough to understand the
 behavior of the fields produced by the isolated source in this region where they are weak.
 This is so because the energy carried by GWs is radiated 
 away and eventually permeates the ``weak-field'' zone. We can extract the power radiated
  by integrating the flux of energy over a
  sphere surrounding the source at a particular time after the emission.
  This calculation is further simplified  in the region far away from the source due to the
    applicability of the both the ``weak-field'' and ``far-zone'' approximations
 (see below).
 
To derive the equations governing the perturbations $h_{\m\n}$ and $\chi$, we split  Eq.~(\ref{eq:eom}) and Eq.~(\ref{eq:eomk}) into linear and non-linear parts as follows
\be
\label{eq:lineq}
\bar G_{\m\n}-M_b^{-2}\bar T_{\m\n}^{\chi}=M_b^{-2}\tau_{\m\n}, \quad
\bar {\mathcal Q}_\chi=-{\mathcal Q}_\chi^{NL}.
\ee
The expression for $\tau_{\m\n}$ reads
\be
\label{eq:tau}
\tau_{\m\n}=T_{\m\n}^m+\left(T_{\m\n}^{\chi}\right)^{NL}-M_b^2G_{\m\n}^{NL}.
\ee
This separation into linear and non-linear parts allows us to solve for $h_{\m\n}$ and $\chi$ perturbatively in $M_b^{-2}$.
 The terms $\tau_{\m\n}$  and ${\mathcal Q}_\chi^{NL}$ can be interpreted to be source terms for the linear equations at different orders in $M_b^{-2}$. They  include
contributions from both matter and non-linear gravitational fields of lower order.  For this paper, we are interested in matter sources that are weakly self-gravitating, 
 slowly moving\footnote{For theories with modes propagating at different speeds, this means
that the typical velocity $v$ of the source is small with respect to all of them.} and weakly stressed. These are known as post-Newtonian (PN) sources 
\cite{Blanchet:2006zz}.  For these systems, one has 
\be
\label{eq:PPNorder}
v\sim |h_{00}^{1/2}|\sim\left| \frac{T^m_{0i}}{T^m_{00}}\right|\sim \left|\frac{T^m_{ij}}{T^m_{00}}\right|^{1/2}\ll 1,
\ee
where $v$ is the typical velocity of the source. Thus, we can introduce $v$ as a new parameter of expansion and consider the predictions of
the theory at different orders in $v$, also known as PN orders. We content ourselves
with the first PN corrections, which amounts to considering Eqs. (\ref{eq:lineq}) where the source terms
also include PN corrections. In particular, the metric should be substituted by its first PN expression (Appendix
\ref{app:PPN}) whenever it appears in the non-linear source terms.

This straightforward analysis is only suited for the first PN corrections. Beyond that, 
the analysis becomes more complicated due to the presence of tails and retardation effects. 
The correct treatment of the problem in general involves the separation into a near-zone and a wave-zone. In the near-zone, 
one can find the metric to any PN  order  including  non-linearities and minimizing retardation effects.
This corresponds to an expansion in the small parameter to desired order in $v$. 
In the wave zone, one can solve the equations of motion perturbatively in the fields and match the solution to the one found in the 
near-zone in a region where both approximations are valid
 \cite{Blanchet:2006zz,Maggiore:1900zz,Blanchet:1989ki,PoissonNotes}. 
For the first PN corrections considered in this paper, this  analysis reduces to the one outlined in the previous paragraph. For
higher order corrections the matching is much less trivial \cite{Blanchet:2006zz,Maggiore:1900zz,Blanchet:1989ki,Pati:2002ux}.

The linearized khronon energy-momentum tensor satisfies the following conservations laws
\be
\label{eq:linearcon}
\pd^\m\bar T_{\m i}^{\chi}=0,\quad  \pd^\m\bar T_{\m 0}^{\chi}=M_b^2\bar{\mathcal Q}_\chi .
\ee
It follows from the invariance of the linearized theory under linear diffeomorphisms.
Next, from  the transverse properties of the $\bar G_{\m\n}$ and 
 when the equations of motion are imposed, one finds
\be
 \pd^\m \left(T_{\m\n}^m+T_{\m\n}^{\chi}-M_b^2G_{\m\n}^{NL}\right)=0.\nonumber
\ee
Together with Eq.~(\ref{eq:linearcon}),  this yields the following conservation equations for the  source tensor $\tau_{\m\n}$  
\be
\label{eq:ncons}
 \pd^\m \tau_{\m\n}=-\pd^\m \bar T_{\m\n}^{\chi}=-M_b^2\delta_\n^0 \,\bar{\mathcal Q}_\chi
\ee
which is of particular importance in Sec. \ref{sec:source} and beyond.

%%%%%%%%%%%%%%%%
\subsection{Wave equations}
%%%%%%%%%%%%%%%%

We decompose the gravitational perturbations
into irreducible representations of $SO(3)$,
\be
\begin{split}
\label{eq:so3}
&h_{00}=2\phi, \quad h_{0i}=-\frac{\pd_i}{\sqrt{\Delta}} B+V_i, \\
&h_{ij}=t_{ij}+2\pd_{(i} F_{j)}+2\frac{\pd_i\pd_j}{{\Delta}}E
+2\left(\delta_{ij}-\frac{\pd_i\pd_j}{{\Delta}}\right)\psi,
\end{split}
\ee
where $t_{ii}=\pd_i t_{ij}=\pd_i V_i=\pd_j F_j=0$.   We also define the Laplacian by $\Delta\equiv \pd_i \pd_i$.

%%%%%%%%%%%%%%%%
\subsubsection*{Tensors and vectors}
%%%%%%%%%%%%%%%%
To single out the tensorial part of the equations of motion as written in Eq.~(\ref{eq:lineq}), we introduce the transverse-traceless projector ${ P}_{ij,kr}$ and the transverse projector $P_{ij}$
\be
\label{eq:proj}
 {P}_{ij,kr}\equiv {P}_{ik}P_{jr}-\frac{1}{2}P_{ij}P_{kr},\quad P_{ij}\equiv \delta_{ij}-\frac{\pd_i \pd_j}{\Delta}.
\ee
A straightforward calculation yields 
\be
{P}_{ij,kr}{\mathcal Q}_{kr}=\frac{1}{2}{ P}_{ij,kr}\left[\beta\, \ddot h_{kr}-(\pd_0^2-\Delta) h_{kr}-2M_b^{-2} \tau_{kr}\right],\nonumber
\ee
leading to the wave equation for the tensor modes
\be
\label{eq:ten}
(c_t^{-2}\pd_0^2- \Delta) t_{ij}=-2M_b^{-2} P_{ij,ks} \tau_{ks},
\ee
with  $c_t^2=1/(1-\beta)$  representing the speed of propagation of the tensor polarizations.
This coincides with the results of \ae-theory \cite{Jacobson:2004ts}. 

Consider now the vectorial part of the equations. Contrary to \ae-theory \cite{Foster:2006az,Jacobson:2004ts} this
sector does not contain any propagating polarizations. Indeed, one finds
\be
\label{eq:vec}
\begin{split}
&P_{ij}\mathcal Q_{0j}=\frac{(1-\beta)}{2}\Delta\left( V_i-\dot F_i\right)-M_b^{-2}P_{ij} \tau_{j0}=0,\\
&P_{ik}\pd_j\mathcal Q_{kj}=\frac{(1-\beta)}{2}\Delta\left(\dot V_i-\ddot F_i\right)
-M_b^{-2} P_{ik}\pd_j \tau_{kj}=0.
\end{split}
\ee
The first equation represents a constraint and its time derivative yields the second equation (which follows from gauge invariance).  For definiteness, we choose to work in the gauge 
\be  \label{eq:gauge1}
F_i=0,
\ee
which completely fixes the gauge freedom in the vector sector.

%%%%%%%%%%%%%%%%
\subsubsection*{Scalars}
%%%%%%%%%%%%%%%%

The scalar sector of the theory is  different from GR. In particular, it includes an extra degree of freedom.
We choose to work in the gauge 
\be \label{eq:gauge2}
\chi=B=0,
\ee
which completely fixes the gauge in the scalar sector.  The choices \eqref{eq:gauge1} and \eqref{eq:gauge2} are referred to as the ``unitary gauge''.
In this gauge, the non-redundant  equations of motion derived from (\ref{eq:eom}) and (\ref{eq:eomk}) are 
\bseq
\label{eq:sca}
\begin{align}
&\alpha\Delta \phi={2\Delta \psi-M_b^{-2}\tau_{00}}, \\
&(\beta+\lambda)\Delta\dot E=-2(\lambda+1)\Delta\dot \psi+M_b^{-2} \pd_i\tau_{0i},\\ \label{eq:scac}
&\left(c_s^{-2} \pd_0^2-\Delta\right)\psi=\frac{\alpha M_b^{-2}}{2(\alpha-2)}\left(\frac{2}{\alpha}\tau_{00}+\tau_{ii}-\frac{(2+\beta+3\lambda)
}{(\beta+\lambda)}\frac{\pd_i\pd_j}{\Delta}\tau_{ij}\right).
\end{align}
\eseq
The speed of propagation of the scalar perturbation is given by  
\be
c_s^2=\frac{(\alpha-2)(\beta+\lambda)}{\alpha (\beta-1)(2+\beta+3\lambda)},
\ee
which coincides with the scalar mode of \ae-theory \cite{Jacobson:2004ts}.

%%%%%%%%%%%%%%%%
\subsection{Far-zone expressions  and post-Newtonian approximation}\label{subsec:farzone}
%%%%%%%%%%%%%%%%

The equations of motion (\ref{eq:ten}), (\ref{eq:vec}) and (\ref{eq:sca}) contain two types of equations that we wish to solve, Poisson and wave equations.  The Poisson equation is of the form
\be
\Delta \xi(t,x)=-4\pi \rho(t,x), \nonumber
\ee
whose solution for vanishing boundary conditions at infinity is given by
\be
\xi(t,x)=\int \di^3 \tilde x\,\frac{\rho(t,\tilde x)}{|x-\tilde x|}.\nonumber
\ee
We assume that the isolated source of GWs can be confined
within a sphere of radius $R$.  At distances far away from the source, $r\equiv |x|\gg R$, 
we can perform the expansion  ($\hat r^i=x^i/r$)
\be
\label{eq:r_exp}
|x-\tilde x|=r- \hat r^i \tilde x^i+ r\, O(R/r)^2.
\ee
We refer to this zone as the ``far-zone''. 
The leading contribution of the solution to the Poisson equation at large distances is then
\be
\label{eq:poiz}
\xi_{f}(t,x)=\frac{1}{r}\int \di^3 \tilde x\, \rho(t,\tilde x).
\ee
The sourced wave equations are of the form
\be
\label{eq:diff}
\left(c_\s^{-2}\pd_0^2-\Delta\right) \sigma(t,x)=4\pi \mu(t,x),
\ee
with speed of propagation $c_\s$. The solution to this equation with  radiation 
boundary conditions is given by (see, e.g. \cite{Maggiore:1900zz})
\be
\label{eq:waveq}
\sigma(t,x)=\int \di^3 \tilde x\,\frac{\mu(t-|x-\tilde x|/c_\s,\tilde x)}{|x-\tilde x|}.
\ee
Besides adopting the far-zone approximation and using (\ref{eq:r_exp}), we also assume that $r$ is such that $r\gg \omega R^2/c_\s$, where $\omega$ is the largest characteristic frequency of the source. 
This allows us to write the leading contribution as 
\be
\label{eq:farz}
\sigma_f(t,x)=
\frac{1}{r}\int \di^3 \tilde x\,{\mu(t-r/c_\s+\hat r^i \tilde x^i/c_\s,\tilde x)}=
\frac{1}{r}\sum_{n=0}^\infty \frac{1}{n! }\pd_0^n\int \di^3 \tilde x{\,\mu(t-r/c_\s,\tilde x)}\left(\hat r^i \tilde x^i/c_\s\right)^n,
\ee 
where the last identity holds formally. 
This expression can be simplified further for the PN sources 
of interest \cite{Will:Book,Blanchet:2006zz,Wein:Book}. 
As seen in the previous section, khronometric theory involves two speeds of propagation, the tensor and the scalar speeds $c_t$ and $c_s$. 
We assume that  the system is slowly moving with respect
to both speeds which are considered to be of the same order,
$c_t\sim c_s\sim 1$.
Thus, for a typical velocity $v\sim \omega R$ of the 
source,
the sum in Eq.~(\ref{eq:farz}) represents a well-defined expansion in the small parameter, 
\be
\nonumber
 v \ll 1,
\ee
i.e. it is a PN expansion, cf. Eq.~(\ref{eq:PPNorder}). In other words, 
every time derivative  in the near zone represents an extra $O(v)$.

%%%%%%%%%%%%%%%%
\section{The source: conservation properties} \label{sec:source}
%%%%%%%%%%%%%%%%

The source terms for the equations  (\ref{eq:ten}), (\ref{eq:vec}) and (\ref{eq:sca}) are expressed in terms of the
pseudo-tensor $\tau_{\m\n}$.  
In order to find solutions to the Poisson and wave equations in the far-zone, Eqs.~(\ref{eq:farz}) and (\ref{eq:poiz}) indicate that we need to evaluate various integrals of $\tau_{\m\n}$.
In what follows, we present results that are relevant
for simplifying those integrals (and therefore the wave forms that appear in Sec.~\ref{sec:waveforms}) and include leading PN corrections.
We refer the reader
to Appendix~\ref{app:PPN} for more details on the first order PN approximation and 
the parametrized-post-Newtonian (PPN) formalism \cite{Will:Book}. 

From Eq.~(\ref{eq:ncons}), one can establish the
 useful integral conservation laws, 
\bseq
\label{integrals}
\begin{align}
&\int\di^3 x\, \tau_{ij}=\frac{1}{2}\int \di^3 x \,\ddot \tau_{00}x^i x^j-\frac{1}{2}\int \di^3 x\, \pd^\m \dot\tau_{\m 0} x^i x^j. \\
&\int\di^3 x\, \dot \tau_{0i} x^j=-\int \di^3 x \,\tau_{ij}.\\
&\int \di^3 x \,\dot \tau_{00}x^i=-\int \di^3 x \, \tau_{i0}+\int \di^3 x \,\pd^\m\tau_{\m0}x^i.\label{integralc}
\end{align}
\eseq

In deriving the previous equations, we assume that all the boundary integrals cancel (the corrections to this assumption are negligible at large $r$).
The difference with respect to the GR integral conservation laws is the presence of the terms proportional to $\pd^{\m} \tau_{\m0}$ coming from the non-conservation of  $\tau_{\m\n}$, Eq.~(\ref{eq:ncons}). Remember  that the current $\tau_{i0}$ is conserved for
khronometric theory .   Naively one expects $\pd^{\m} \tau_{\m0}$ to contribute to order as low as $O(v^3)$.  To see that this is not the case, we notice that Eqs. (\ref{integrals}) can be simplified  by writing the equation of motion (\ref{eq:eomk})  as an equation for a conserved current (which corresponds to the Noether current related to the invariance of the theory under reparametrizations of $\varphi$, Eq.~(\ref{eq:repr})), 
\be
\label{eq:consc}
\pd_\m(\sqrt{-g}J^\m)=0.
\ee
Furthermore, in the unitary gauge, $\varphi=t$ and $J^0=0$ (see Appendix D of \cite{Blas:2010hb}).
Since $J^i$ is linear in perturbations, we find that
\be
\label{eq:chiNL}
\bar{\mathcal Q}_\chi=-\pd_i (\sqrt{-g}J^i)^{NL}.
\ee
Thus,
\be
\label{eq:sigma}
\pd^\m\tau_{0\m}=
M_b^2 \pd_i(\sqrt{-g}J^i)^{NL}\sim O(v^5),
\ee
and the dipolar corrections turn out to be large in PN order. 
In particular,  at order $O(v^5)$ only the Eq. (\ref{integralc}) is modified with respect to GR.
A straightforward but tedious calculation using the PN metric displayed in  Eq.~\eqref{eq:PNunitg} yields
\be
\int \di^3 x \,\pd^\m\tau_{0\m}x^i=\frac{1}{2}\int \di
 x \ \rho\left[(\alpha^{PPN}_1-\alpha^{PPN}_2)V^{PPN}_i+\alpha_2^{PPN} W^{PPN}_i\right]+O(v^6),\label{eq:dipoleU}
\ee
where
\be
\label{eq:alph}
\begin{split}
 \alpha^{PPN}_1&=\frac{4(\alpha-2\beta)}{\beta-1}, \\
\alpha^{PPN}_2&=\frac{(\alpha-2\beta)(-\beta[3+\beta+3\lambda]-\lambda+\alpha[1+\beta+2\lambda])}{(\alpha-2)(\beta-1)(\beta+\lambda)}.
 \end{split}
 \ee
 These constants are the PPN parameters related to the violation of Lorentz invariance of the theory (see Appendix~\ref{app:PPN}).
 In the limit of small parameters they coincide with those found in \cite{Blas:2010hb}.
 The potentials  $V^{PPN}_i$ and $W^{PPN}_i$ are also
defined in  Appendix~\ref{app:PPN}. Finally, the form of the Eq. (\ref{eq:dipoleU}) is identical to the one found for \ae-theory \cite{Foster:2007gr}.

The previous formulae \eqref{integrals} and \eqref{eq:dipoleU} can also be derived by relating the pseudo-tensor $\tau_{\m\n}$ to a conserved (but asymmetric) object.
Indeed, from Eqs.~(\ref{eq:ncons}) and (\ref{eq:eomk}) it is evident that
\be
\mathfrak{T}_{\m\n}\equiv\tau_{\m\n}+M_b^2\delta_\m^0\,\eta_{\n\r}\bar J^{\r}\nonumber
\ee
satisfies
\be
\pd^\n \mathfrak{T}_{\m\n}=0.\nonumber
\ee
This object has a contribution linear in the fields. To build a quadratic conserved pseudo-tensor it is enough to add
the conserved current found in (\ref{eq:consc}) and consider the object 
\be
\mathfrak{T}_{\m\n}^q\equiv \tau_{\m\n}+M_b^2\delta_\m^0\eta_{\n\r}(\bar J^{\r}-\sqrt{-g}J^\r)
=\tau_{\m\n}-M_b^2\delta_\m^0\,\eta_{\n\r}(\sqrt{-g}J^\r)^{NL}.\nonumber
\ee
The resulting integral conservation laws for this object are then identical to the ones found in \cite{Will:Book,Foster:2007gr}.  The existence of this conserved quadratic current is a generic consequence of the theory being \emph{semi-conservative} in the
language of \cite{Will:Book}. This conserved current satisfies
\be
\mathfrak{T}^q_{0i}-\mathfrak{T}^q_{i0}=M_b^2(\sqrt{-g}J^i)^{NL}.\nonumber
\ee
Then, one can  use the  Eq.~(4.103) in \cite{Will:Book} to compute (\ref{eq:sigma}). Even if this 
method may save a lot of computations, it is inconvenient since the result in 
\cite{Will:Book} is derived
in the PPN gauge, whereas we are interested in the result in the unitary gauge (\ref{eq:dipoleU}).

%%%%%%%%%%%%%%%%
\section{Wave forms in the far-zone}\label{sec:waveforms}
%%%%%%%%%%%%%%%%

We are now ready to compute the explicit form of the wave solutions in the far-zone,  which we do consistently up to $O(v^6)$ in the PN approximation.
 For the tensor and vector modes,
the solutions of Eqs.~(\ref{eq:ten}), (\ref{eq:vec}), (\ref{eq:farz}) and (\ref{integrals}) are (in the gauge $F_i=0$)
\bseq
\begin{align}
&t_{ij}(t,x)=-\frac{1}{4\pi M_b^2 r} \hat P_{ij,ks} \ddot Q_{ks}(t-r/c_t)-\frac{1}{2\pi M_b^2c_t\,r} \hat P_{ij,ks}
\hat r^a\dot S_{ks,a}(t-r/c_t)+O(v^6),\label{eq:wtensor}\\
&V_i(t,x)=-\frac{c_t^2}{2\pi M_b^2}\left(\frac{1}{r}\int\di^3 \tilde x\ P_{ij} \tau_{j0}(t,\tilde x)\right),\label{eq:wvector}
\end{align}
\eseq
where
\be
\begin{split}
Q(t)_{ij}\equiv I(t)_{ij}-\frac{1}{3}\delta_{ij} I_{kk}(t)&, \quad I_{ij}(t)\equiv\int \di \tilde x\, \tau_{00}(t,\tilde x) \tilde x^i\tilde x^j,\\
\quad S_{ks,a}(t)&\equiv\int \di^3 \tilde x\ \tau_{ks}(t,\tilde x)\tilde x^a.\nonumber
\end{split}
\ee
The quantity  $Q_{ij}$ represents the quadrupole of $\tau_{00}$.  Note that in the far-zone, the longitudinal projector  $P_{ij}$ of Eq.~(\ref{eq:proj}) can be substituted by the longitudinal part of the algebraic projector,
\be
\hat P_{ij}\equiv \delta_{ij}-\hat r^i \hat r^j.\nonumber
\ee
This substitution is valid up to $O(R/r)$ terms. The  object $\hat P_{ij,ks}$ is defined as
 $${\hat P}_{ij,kr}\equiv \hat P_{ik}\hat P_{jr}-\frac{1}{2}\hat P_{ij}\hat P_{kr}.$$
Anticipating the results of Sec.~\ref{sec:energy},
we notice that the energy-loss formula depends on the time derivative of the fields.  For the vector
part, the previous expression yields
\be
\label{eq:dotV}
\dot h_{0i}=\dot V_i=-\frac{c_t^2}{2\pi M_b^2}\left(\frac{1}{r}\int\di^3 \tilde x\ P_{ij} \dot\tau_{j0}(t,\tilde x)\right).
\ee
From the conservation law (\ref{eq:ncons}), this term can be expressed as the integration over the boundary of the
transverse component of the source, which cancels away from the source, and we can neglect the vector
perturbations altogether.

Concerning the scalar field,
from the wave equation (\ref{eq:scac}) one finds
\be
\label{eq:scawave}
\begin{split}
\psi=
&\frac{\a}{8\pi(\a-2)M_b^2r}\Bigg(
\frac{3}{2}\left[Z-1\right]\hat r^i \hat r^j \ddot Q_{ij}(t-r/c_s)+
\frac{1}{2}Z\, \ddot I_{kk}(t-r/c_s)
\\
&+\frac{2}{c_s\a}\hat r^i\int \di^3 \tilde x\ {\dot\tau_{00}(t-r/c_s,\tilde x)}\tilde x^i
+\frac{1}{3c_s^3\a}\hat r^i \hat r^j \hat r^k \int \di^3 \tilde x\dddot \tau_{00}(t-r/c_s,\tilde x)\tilde x^i\tilde x^j\tilde x^k
\\&+\frac{1}{c_s}\hat r^a \dot S_{kk,a}(t-r/c_s)-\frac{(2+\beta+3\l)}{c_s(\b+\l)}\hat r^i\hat r^j \hat r^k \dot
S_{ij,k}(t-r/c_s) \Bigg)+O(v^6),
\end{split}
\ee
where
\be
Z\equiv\frac{(\beta-1)(\a^{PPN}_1-2\a^{PPN}_2)}{3(\a-2\b)}.\label{eq:Z}
\ee
Notice that the conservation law $\dot \tau_{00}=\pd_i(\tau_{0i}+\bar J_i)$ has been used to show that the first moment
of $\tau_{00}$ is constant in time and therefore ignored in \eqref{eq:scawave}.
From the results in the previous section, we see that the modification to the GR
results appear at order $O(v^4)$.  Notice also that it follows from (\ref{eq:dipoleU}) and the
constancy of the integral of $\tau_{i0}$ that
 the dipolar contribution in (\ref{eq:scawave}) is $O(v^5)$ and suppressed by
the PPN parameters. Finally, the quadrupole terms in the tensor and scalar sectors differ slightly, as they depend on different retarded times.
For the remaining scalar fields $\phi$ and $E$, from Eqs. (\ref{eq:sca}) one finds
\be
\label{eq:scalarsol}
\phi=\frac{2}{\alpha}\psi+\frac{1}{4\pi M_b^2\alpha\, r}\int \di^3 \tilde x\, \tau_{00}, \quad \dot E=-\frac{2(\lambda+1)}{\beta+\lambda}\dot \psi.
\ee

%%%%%%%%%%%%%%%%
\section{Energy-loss formulae for post-Newtonian systems}\label{sec:energy}
%%%%%%%%%%%%%%%%

The definition of the energy carried by gravity waves is non-trivial (see \cite{Szabados:2004vb} for a review on the
 concepts of energy and momentum in GR). 
For the problem at hand, we follow the procedure of \cite{Foster:2007gr} (see also \cite{Foster:2005fr}) 
and use the
notion of energy for asymptotically flat spacetimes derived in \cite{Iyer:1994ys}. Given an isolated source, we can 
compute the time variation of this notion of energy by performing an integral of the flux in the far-zone, which we idealize as being
infinitely far away from the source. We associate this energy loss to the energy carried away by gravitational radiation.
As shown in \cite{Iyer:1994ys,Sorkin:1991bw}, this alternative approach is equivalent to the one
based on pseudo-tensors used in standard computations
 of energy loss due to gravitational radiation 
\cite{Will:Book,Maggiore:1900zz,Wein:Book}. We give a brief review on this method in Appendix~\ref{ap:energy},

In deriving the energy-loss formula, we make the following assumptions.  We start by assuming that our system consists of an asymptotically Minkowski spacetime at early times,  with the following fall-off properties in the unitary gauge,
\be
\label{eq:fallPPN}
g_{\m\n}=\eta_{\m\n}+O(1/r), \quad \pd_\a g_{\m\n}=O(1/r^2), \quad \chi=0.
\ee
As for the matter fields, we assume that they vanish asymptotically to ensure that there are no boundary integral contributions.
The previous conditions allow us to define
a convenient notion of conserved energy $\mathcal E$, Eq.~(\ref{eq:energyf}), as the conserved charge associated to the invariance of the asymptotic solution under asymptotic time translations\footnote{Even if this symmetry is broken by the background for the field $\varphi$,
it is still a symmetry due to the reparametrization invariance of the theory (\ref{eq:repr}). }.
 To compute the flux of gravitational radiation, we consider the moment of time when the emitted GWs are
 already at spatial infinity, which means that the 
 fall-off properties of the fields change to
 \be
 \label{eq:boundw}
 h_w\sim O(1/r), \quad \dot h_w\sim \omega O(1/r), \quad \pd_r h_w\sim \omega/c_s O(1/r).
 \ee
The quantity $\mathcal E$  with these boundary conditions is in general divergent.
Nevertheless,  its change  due to the radiation emitted
 during a finite interval of time is well-defined \cite{Regge:1974zd}.  We focus on computing the time variation of $\mathcal  E$, 
 Eq.~\eqref{eq:energyfl}. As shown in Appendix~\ref{ap:energy},  $\dot{\mathcal E}$ is finite
for conditions (\ref{eq:boundw}) and only has contributions that are quadratic in the fields.

We also consider the time average of the quantity $\dot{\mathcal E}$  over several periods of the source,
$\langle \dot{\mathcal  E} \rangle.$
 The final averaged energy-loss formula is a standard observable in GW experiments (including the binary system of interest, where the
 observed damping of the orbits occurs after several periods) and the final expression is simplified since total time derivatives vanish when integrated. 
  
The final expression is further simplified after one takes into account the following considerations. 
From the form of the solution of the tensor modes,  Eq.~(\ref{eq:wtensor}), and of the field $\psi$, Eq.~(\ref{eq:scawave}), 
 then in the far-zone these fields satisfy the equation
\be
\label{eq:waverel}
c_\s\,\pd_i \sigma=-\hat r^i\dot\sigma,
\ee
for the corresponding speeds of propagation.
Remember also that in the far-zone, the tensor modes $t_{ij}$ are transverse with respect to the algebraic projector  $\hat r^i t_{ij}=0$.
For the vector part,
we already showed that it does not contribute to $ \langle \dot{\mathcal  E} \rangle$, as its time derivative
cancels, cf. Eq.~(\ref{eq:dotV}). Similarly, the fields $E$ and $\phi$ always appear under a time or a space derivative.
Thus, we notice that the dependence on the source  appearing in Eq. (\ref{eq:scalarsol}) will be either higher order in
$R/r$ for the space derivatives (and therefore negligible), or of the form
\be
\int \di^3 \tilde x\, \dot\tau_{00}=-M_b^2\int \di^3 \tilde x\, \bar {\mathcal Q}_\chi=
M_b^2\int \di^3 \tilde x\, \pd_i (\sqrt{-g}J^i)^{NL}=0.\nonumber
\ee
So, only the $\psi$ contribution for $E$ and $\phi$ is non-zero, and therefore these fields satisfy relation (\ref{eq:waverel}). In fact, 
the latter relation is also satisfied by the scalar part of $h_{ij}$.   

The previous considerations (and some algebra presented in the Appendix~\ref{ap:energy}) yield the final result 
for the rate of energy loss of the system, 
\be
\label{eq:dotener}
\langle \dot{\mathcal  E} \rangle=-\frac{M_b^2}{4}\oint_{S_\infty^2}  \di \Omega \,  r^2\left
\langle \frac{1}{c_t}\dot t_{ij} \dot t_{ij}-\frac{8(\alpha-2)}{\alpha\,c_s}\dot \psi \dot \psi\right\rangle.
\ee
Whereas the radiation emitted in the tensor modes always decreases the energy of the system, the behaviour of the emitted scalar modes depends
on the parameter $\alpha$. We see that the emitted energy is positive for the range $0<\alpha< 2$, as expected
since these values are also required for the stability of the Minkowski background (absence of ghosts) \cite{Blas:2009qj}.

Up to this point, we have consistently worked to first PN order (which corresponds to and includes $O(v^5)$ in the wave-forms). 
Given the time derivatives in Eq. (\ref{eq:dotener}), substitution of the waveforms (\ref{eq:wtensor}) and (\ref{eq:scawave}) yields
the energy loss of the system from gravitational radiation up to and including  $O(v^{12})$, although corrections already appear at leading order, $O(v^{10})$.  To simplify what follows, we therefore focus on Newtonian
sources and the corrections  at this order. 
Substituting the waveforms to lowest PN order in the previous expressions and performing the angular integrals, 
we find the energy-loss formula\footnote{In this final formula, we compute the quadrupole and monopole terms at a time when 
 radiation from both the tensor and scalar modes simultaneously reaches the boundary of the isolated system. 
For different speeds of propagation $c_t$ and $c_s$, the discrepancy in emission time is irrelevant for stationary production of GWs.}
\be
\label{eq:dotenerN}
\langle \dot{\mathcal  E} \rangle=-\frac{1}{8\pi M_b^2}\left\langle \frac{\mathcal A}{5}\dddot Q_{ij}\dddot Q_{ij}+\mathcal B
\dddot I\dddot I \right\rangle,
\ee
where (recall Eq.~(\ref{eq:Z}))
\be
\mathcal A\equiv \frac{1}{c_t}-\frac{3\a(Z-1)^2}{2c_s(\alpha-2)}, \quad \mathcal B=-\frac{\alpha Z^2}{4c_s(\alpha-2)}.\nonumber
\ee
The final expression, Eq.~\eqref{eq:dotenerN}, differs from the GR result in two ways: the coefficient corresponding
to the quadrupole depends on the parameters of the model;  there is a monopole contribution already at this 
first Newtonian order. This is similar to the formula derived for \ae-theory\footnote{\label{FosterM}For the monopole and quadrupole contributions, the definition of $Z$ in \cite{Foster:2007gr} differs from Eq.~(\ref{eq:Z}) by a factor of two.  We attribute this difference to a typo in the final formula for $\psi$ in \cite{Foster:2007gr}.} \cite{Foster:2007gr}.
Let us note something quite remarkable in the context of the khronometric theory that we are studying.   All of the Solar System tests 
are passed in the limit $|\alpha_1^{PPN}|\ll 1$, $|\alpha_2|^{PPN}\ll 1$, which can be achieved by the single requirement
$\vert \alpha-2\beta \vert \ll 1$, cf. (\ref{eq:alph}). In this limit, $Z=1$, the dipole term (\ref{eq:dipoleU}) cancels and the
 monopole contribution in Eq. (\ref{eq:dotenerN}) is still present. This last result contrasts with the \ae-theory case for which there is only a modified quadrupole (in the equivalent limit). This discontinuity between the two theories is discussed in Appendix \ref{app:monopole}.

%%%%%%%%%%%%%%%%
\section{Energy loss by a Newtonian binary system}\label{sec:binary}
%%%%%%%%%%%%%%%%

To complete the calculation, the power-loss formula (\ref{eq:dotenerN}) must be supplemented by the equations of motion
of the system to desired post-Newtonian (PN) order. We content ourselves with a  2-body Newtonian system composed of point-masses $m_1$ and $m_2$.  The matter action is then given by
\be
\label{eq:matter}
S^m=-\sum_{A=1}^2 \int m_A  \di s_A \,,
\ee
where $ \di s_A$ represents the proper-time of the $A$-th particle.
A priori,  $m_A$ depends on the khronon field.  Since we are only interested in the Newtonian system, it is enough to Taylor expand the mass around its background value and use only the leading order contribution, hence $m_A$ is taken to be constant.  Using the preferred time as the affine parameter,  the energy-momentum tensor derived from (\ref{eq:matter}) is
\be
T^m_{\m\n}=\frac{1}{\sqrt{-g}}\sum_{A=1}^2\frac{m_A u_{A\m} u_{A\n}}{\sqrt{g_{\r\s} u_A^\r u_A^\s}}\delta^{(3)}(x^k-x_A^k(t)),\nonumber
\ee
where the $A$-th body follows the trajectory $x_A^k(t)$ with four-velocity $u_A^\m$. At Newtonian order this yields
\be
\ddot I_{ij}(t-r/c_\s)=\pd_0^2\sum_{A=1}^2 m_A x_A^i(t-r/c_\s)x_A^j(t-r/c_\s).\nonumber
\ee
We evaluate this at very late times as explained in the previous section.
Next, from the geodesic equation derived from (\ref{eq:matter}), we find Newton's law
$$
\ddot x^i_1= -G_N\frac{m_2}{r_{12}^2}\hat r_{12}^i, \quad \ddot x^i_2= G_N\frac{m_1}{r_{12}^2}\hat r_{12}^i,
$$
where we introduce (also see Appendix~\ref{app:PPN})
 \be
 G_N\equiv\frac{1}{4\pi M_b^2(2-\a)}, \quad  r_{12}^i\equiv x^i_1-x_2^i.\label{eq:GN}
\ee
As usual in binary systems, it is convenient to define the problem in terms of the relative distances and the position of the
center of mass  $x_{CM}\equiv \frac{m_1 x_1+m_2 x_2}{m_1+m_2}$.
Finally, assuming\footnote{ Corrections to this assumption are considered as higher order in the PN expansion.} that the system is at rest with respect to the preferred frame (so that $\dot x_{CM}=0$) we get
\be
\begin{split}
\dddot I_{ij}(t-r/c)=-\frac{2G_N \m M}{r_{12}^2}\left( 4 \hat r_{12}^{(i} v^{j)}-3 \hat r_{12}^{i} \hat r_{12}^{j} \dot r_{12} \right).\nonumber
\end{split}
\ee
with $\m\equiv m_1m_2/M$, $M\equiv m_1+m_2$ and $v^i \equiv \dot{r}_{12}^i $ is related to the expansion parameter $v$.
Thus, the loss of energy in gravitational radiation for a Newtonian binary system is given by
\be
\label{eq:edotNB}
\langle \dot{\mathcal  E} \rangle=-\frac{1}{\pi M_b^2}\left(\frac{G_N M\m }{r_{12}^2}\right)^2\left\langle \frac{1}{15}\mathcal A
\left(12 v^2-11 \dot r_{12}^2\right)+\frac{\mathcal B}{2}\dot r_{12}^2\right\rangle,
\ee
from which we deduce the `Peters-Mathews' (PM) parameters  \cite{Will:Book,Peters:1963ux} ($\kappa_D=0$),
\be
\kappa_1=12(1-\alpha/2)\mathcal A,\quad \kappa_2=(1-\alpha/2)\left(11\mathcal A-\frac{15}{2}\mathcal B\right).\nonumber
\ee
Once the energy loss for the binary system is known, one can use Kepler's third law
to relate it to the damping of the orbit. The expression for the
change of the orbit's period for generic PM parameters
 in terms of other orbital parameters of the system can be found in \cite{Weisberg:1981bh}.

In GR, the previous analysis suffices to predict the radiation damping  of binary systems for compact (relativistic) sources, like the PSR1913+16 \cite{Will:Book}. This is because the structure of the compact stars of the binary does not
influence the orbit  in GR (this is called the `effacing principle' which is a consequence of the strong equivalence principle). This is certainly not true
for most alternative theories of gravity. Thus, to yield concrete predictions about the radiation
damping of systems with highly relativistic sources (sources with large self-energies), one must first understand the behaviour of the
fields beyond the PN approximation. One can then use Eq.~(\ref{eq:dotenerN}) to derive the energy loss resulting in a change of the orbit (at corresponding PN order). For scalar-tensor theories, the final result is a  test
of the strong-field regime \cite{Eardley75,Damour:1996ke,EspositoFarese:2004cc}.
For \ae-theory, the first steps were performed in \cite{Foster:2007gr}
based on the stellar solutions of \cite{Eling:2007xh} and using the effective field
theory methods of \cite{Goldberger:2004jt,Porto:2006bt,Porto:2005ac,Goldberger:2009qd}

It is beyond the scope of our article to derive the radiation damping of
these realistic systems (including relativistic self-gravitating objects) for khronometric
theory. In any case, we do not expect the new corrections to cancel the ones we have already
derived for the Newtonian source, and thus we find it appropriate to use Eq.~(\ref{eq:edotNB}) to
set order of magnitude bounds on the free parameters
of the khronometric action (\ref{covar22}). Current data on the radiation damping of the Hulse-Taylor binary system agrees with GR
up to a level slightly better than one part in one hundred \cite{Will:2005va,Weisberg:2004hi}. This means
that the formula (\ref{eq:edotNB}) should agree with GR to $O(10^{-2})$, which finally implies the
bound (for the case where $\a$, $\b$ and $\l$ are of the same order)
\be
\label{eq:GWbound}
\alpha \sim \b \sim \l \lesssim 10^{-2}.
\ee
The previous bound is less stringent than  the  bounds coming from the PPN
analysis  \cite{Blas:2010hb,Will:Book}, $$ |\a_1^{PPN}|\lesssim 10^{-4},\quad~|\a_2^{PPN}|\lesssim~10^{-7}.\;$$
As can be directly seen from Eq.~(\ref{eq:alph}), the PPN  bounds are automatically satisfied in the limit $\alpha=2\beta$.
In this limit, $Z=1$, and our expression (\ref{eq:GWbound}) yields the most stringent bound for the theory.
Notice in particular that it constrains the propagation speeds to be close to $c=1$.
Another constraint in this limit comes from the difference between $G_N$ as
derived in (\ref{eq:GN}) and the value for  Newton's constant appearing in Friedmann's equation, $G_c$ \cite{Blas:2009qj}. The value of $G_c$
 is constrained by nucleosynthesis and satisfies $\left|\frac{G_N}{G_c}-1\right|\leq 0.13$ \cite{Carroll:2004ai}, 
 which, in terms of the parameters in the action (\ref{covar22}), 
 implies the estimate $\alpha,\, \beta,\, \lambda \lesssim 0.1$ \cite{Blas:2009qj}. Also, we should ensure the absence of gravitational Cherenkov  radiation, which implies  $c_t^2\geq 1$ and $c_s^2\geq 1$ (this means that a particle moving through the aether does not radiate\footnote{We thank D. Levkov and S. Sibiryakov
for pointing this out to us.}).   Notice that the speeds are superluminal,
which does not pose a threat to Lorentz violating theories as long as causality is maintained.

%%%%%%%%%%%%%%%%

\section{Discussion}\label{sec:discussion}
%%%%%%%%%%%%%%%%

Our aim has been to study the radiation loss from an isolated 
source in the PN approximation for khronometric theory. This theory is an interesting alternative to GR with a high energy cutoff and
for which
a UV completion is known in the form of Ho\v rava gravity.  It is also very similar to \ae-theory, 
as in both cases there is a preferred time coordinate. The difference is that khronometric theory has only one extra scalar degree
of freedom, the khronon, whereas \ae-theory relies on a timelike unit dynamical vector leading to three extra degrees of freedom, consisting of one scalar and one vector field.

For arbitrary parameters, we have shown in Eq.~(\ref{eq:dotenerN}) that the formula controlling the power loss of the system (which 
may be related to the change of the orbital period of a binary source) is modified with respect to GR already at lowest, Newtonian, order.
 In particular, the quadrupole contribution differs from GR, partly due to the different
 speeds of propagation of the tensor modes in both theories.
Furthermore, there is also an extra monopole contribution at this order.  The monopole at leading order in khronometric theory contrasts with the usual situation in other scalar-tensor theories \cite{Damour:1992we}.  At higher order, there are other modifications, including
the dipole term (\ref{eq:dipoleU}).  Quite remarkably, in the phenomenologically 
interesting limit where all PPN parameters coincide with GR (which corresponds to the limit  $\a=2\b$
for our parameters), the monopole is still present, and its strength is proportional to the 
parameters appearing in the action of the theory, Eq.~(\ref{covar22}). 
These results for khronometric theory are similar to those of \ae-theory, modulo vector
propagating degrees of freedom that are absent for the khronometric case. There is a key difference, however, since \ae-theory only has a modified quadrupole to lowest order in the equivalent limit.  

This work has been devoted to PN sources. These types of sources do not correspond to the ones found in the binary systems of interest, 
which are compact and characterized by strong gravitational fields. Despite this, we have
 evaluated the energy-loss formula for the simplest possible system: a Newtonian binary.
Doing so provides an order of magnitude estimate on the parameters of the theory (as we 
do not expect corrections due to strong-fields to cancel the modifications apparent in the power loss formula). Thus, our results are relevant
for constraining the case $\a=2\beta$. In this case, requiring the rate of radiation damping to be close to GR sets constraints on this parameters of order $O(10^{-2})$.  These constraints represent the strongest phenomenological bounds for this particular choice of
parameters and are relevant for the 
cosmological implications of the theory, including the recently suggested model of dark-energy \cite{Blas:2011en}.

Sources with strong self-energies is left for future research and can be treated in our theory in the same way as scalar-tensor theories \cite{Eardley75,Damour:1996ke,EspositoFarese:2004cc}: a phenomenon of ``scalarization''  modifies the orbit of these sources as compared
with the post-Newtonian ones.   It would also be interesting to consider our results in the 
parametrized post-Einsteinian framework introduced in \cite{Yunes:2009ke} (see also  \cite{Yunes:2010qb} for the
 binary pulsar
constraints for this framework). Finally, the consequences 
of alternatives theories of gravity for experiments of direct detection of GWs  have been recently discussed, see e.g. 
\cite{Will:2005va,DelPozzo:2011pg}. We hope to extend these works to khronometric theories in the future.

%%%%%%%%%%%%%%%%%
\subsection*{Acknowledgments}
%%%%%%%%%%%%%%%%%

We would like to thank Brendan Foster, Ted Jacobson, Rafael Porto, Oriol Pujol\`as and Thomas Sotiriou for useful discussions.  We are 
grateful to Gilles Esposito-Far\` ese, Michele Maggiore and Riccardo Sturani for their helpful comments on the draft. We are especially indebted to Sergey Sibiryakov for illuminating conversations and invaluable comments on the manuscript. The work of D.B. and H.S. is funded by the Swiss National Science Foundation.  
\appendix

%%%%%%%%%%%%%%%%
\section{Post-Newtonian expressions}\label{app:PPN}
%%%%%%%%%%%%%%%%

The PPN formalism is a valuable tool for comparing theories of gravitation with each other and with experiment
in the weak, non-relativistic limit \cite{Will:Book} .
  In this section, we  briefly present the steps involved in the PPN calculation for khronometric theory 
  (see also \cite{Blas:2010hb}). The final result 
  are the parameters (all the other PPN parameters cancel)
\be \label{eq:PPN}
\begin{split}
\beta^{PPN}&=\gamma^{PPN}=1, \\
\alpha^{PPN}_1&=\frac{4(\alpha-2\beta)}{\beta-1}, \\
\alpha^{PPN}_2&=\frac{(\alpha-2\beta)(-\beta(3+\beta+3\l)-\lambda+\alpha[1+\beta+2\lambda])}{(\alpha-2)(\beta-1)(\beta+\lambda)}.
\end{split}
\ee
Notice that the PPN parameters for
  khronometric theory for arbitrary values of the parameters in (\ref{covar22}) appear here for the first time. 
  They coincide with  results in \cite{Blas:2010hb} in the limit of small parameters.
The non-zero parameters $\alpha^{PPN}_1$ and $\alpha^{PPN}_2$ indicate that khronometric theory violates Lorentz invariance.  These same two parameters are non-vanishing 
 for \ae-theory, although the dependence on the parameters $\a$, $\b$ and $\l$ is different.  In both theories, however, the relationship between  $\alpha^{PPN}_1$ and $\alpha^{PPN}_2$ is the same
 \be \nonumber
\alpha^{PPN}_2=\frac{\alpha^{PPN}_1}{2}-\frac{ (2 \b- \a) (3 \lambda + \b + \a)}{(\lambda+\b) (2-\a)}.
\ee

To compute the previous results we closely follow  \cite{Will:Book,Foster:2005dk} to which we refer the reader for further details.
The source is assumed to be a fluid with a covariantly conserved energy-momentum tensor
\be \nonumber
T^{\mu \nu}=(\rho+\rho \Pi+p)v^{\mu}v^{\nu}- pg^{\mu\nu},
\ee
where $v^{\mu}$ is the four velocity of the source, $\rho$ the rest mass energy density, $\Pi$ the internal energy density and $p$ the isotropic pressure of the fluid.   The source is assumed to satisfy (\ref{eq:PPNorder}).

In what follows, recall that the different fields  have the following expansion,
\be
\label{eq:PNo}
\begin{split}
&g_{00}=1+O(v^2)+O(v^4), \quad g_{0i}=O(v^3), \\
& g_{ij}=-\delta_{ij}+O(v^2), \quad \chi= O(v^2)+O(v^3).
\end{split}
\ee
 Also, we use the following potentials
\be  \nonumber
    F(x) =G_{\rm N} \int d^3y \frac{\rho(y)f}{|x-y|},
\ee
where $G_{\rm N}$  is defined in Eq.~(\ref{eq:GN}) and the
correspondence $F\mapsto f$ is given by
\begin{gather}  \nonumber
    U \mapsto 1,\quad
    \Phi_1 \mapsto v_i v_i,\quad
    \Phi_2 \mapsto U,\quad
    \Phi_3 \mapsto \Pi,\quad
    \Phi_4 \mapsto p/\rho, \\
    V_i^{PPN} \mapsto v^i,\quad
    W_i ^{PPN} \mapsto \frac{v_j (x_j - y_j)(x^i -y^i)}{|x-y|^2}\nonumber.
\end{gather}
The steps to take are: 
\begin{enumerate}
\item Solve $g_{00}$ to order $O(v^2)$.  For this we use the $00$ component of  Eq.~(\ref{eq:eom})
to $O(v^2)$, which yields\footnote{We use a number over the field to keep track of the order in $v$.}
$$
\Delta  \overset{2}h_{00}=8\pi G_N \rho.
$$
\item {Solve $g_{ij}$ to $O(v^2)$.}  Following \cite{Foster:2005dk}, we choose the gauge conditions
\be \nonumber
 \pd_i\overset{2}h_{ij}=-\frac{1}{2}\left(\pd_i\overset{2}h_{00}-\pd_i \overset{2}h_{kk}\right),
\quad \pd_i \overset{3}h_{0i}=\Gamma\pd_0\overset{2}h_{00}.
\ee
The arbitrary constant $\Gamma$ will be chosen to write the result in the PPN gauge.
Then from the $ij$ component of Eq.~(\ref{eq:eom}) to $O(v^2)$, we find
\be \nonumber
\Delta \overset{2}h_{ij}=8\pi G_N\rho\, \delta_{ij}.
\ee
\item {Solve $\chi$ to $O(v^3)$.}  The khronon equation of motion  \eqref{eq:eomk} to leading order is given by 
\be \nonumber
\left(\,\Delta \overset{3}\chi-\Gamma \pd_0 \overset{2}h_{00}\right)=
-\frac{(3\lambda+\alpha+\beta)}{2(\lambda+\beta)}\pd_0 \overset{2}h_{00},
\ee
\item {Solve $g_{0i}$ to $O(v^3)$.} 
In our gauge, the $0i$ component of Eq.~(\ref{eq:eom}) to $O(v^3)$ yields
\be \nonumber
\begin{split}
\Delta \overset{3}h_{0i}=\frac{8\pi G_N \rho\, v_i(\alpha-2)-[-2+\alpha+\Gamma(1-\beta)]\pd_0\pd_i\overset{2}h_{00}}{\beta-1}.
\end{split}
\ee
\item  {Solve $g_{00}$ to $O(v^4)$.} 
From  the $00$ component of Eq.~(\ref{eq:eom}) to $O(v^4)$, we find
\be \nonumber
\begin{split}
&\Delta \overset{4}h_{00}=\pd_i \overset{2} h_{00}\pd_i \overset{2} h_{00}-
\overset{2} h_{00}\Delta\overset{2} h_{00}-4 \Delta\Phi_1+4 \Delta\Phi_2-2 \Delta\Phi_3-6 \Delta\Phi_4\\
&+\frac{(-\alpha^2+2\beta(3+\beta-2\Gamma)+2(3+3\beta-2\Gamma)\lambda+2\alpha(\beta(\Gamma-1)
+(\Gamma-3)\lambda))}{(\alpha-2)(\beta+\lambda)}\Delta \pd^2_0 H,
\end{split}
\ee
where $H=-G_N\int \di^3 y \rho |x-y|$ is known as the superpotential.

\item To go to the PPN gauge, we choose $\Gamma$ that cancels the term depending on $H$ in the previous
equation \cite{Will:Book}.
\end{enumerate}

Putting everything together, we have (to desired order)
\be \nonumber
\begin{split}
g^{PPN}_{00}&=1-2U+2U^2-4\Phi_1-4\Phi_2 -2\Phi_3-6\Phi_4=1+\Delta H,\\
g^{PPN}_{ij}&=-(1+2U)\delta_{ij}=\delta_{ij}\left(-1+\Delta H\right), \\
g^{PPN}_{0i}&=\frac{1}{2}(7+\alpha_1^{PPN}-\alpha_2^{PPN})V_i^{PPN} +\frac{1}{2}(1+\alpha_2^{PPN})W_i^{PPN},\\
\chi^{PPN}&=\frac{(\alpha-2\beta)(2+\beta+3\lambda)\dot H}{2(\alpha-2)(\beta+\lambda)},
\end{split}
\ee
which, compared to the generic PPN metric  (see for example, Eq.~(A.11) of \cite{Foster:2005dk}) implies that all the PPN parameters vanish except for the ones cited in \eqref{eq:PPN}.  

The PN metric in the unitary gauge of Eqs.~\eqref{eq:gauge1} and \eqref{eq:gauge2} is easily derived from these PPN expressions.  It suffices to go from the PPN gauge to the unitary gauge via a diffeomorphism $\delta x^\m=\xi^\m$ satisfying
\be  \nonumber
\begin{split}
&\xi^0=-\frac{(\alpha-2\beta)(2+\beta+3\lambda)\dot H}{2(\alpha-2)(\beta+\lambda)},\quad
\xi^i=\frac{(\alpha+\beta+3\lambda)\pd_i H}{2(\beta+\lambda)}.
\end{split}
\ee
This leads to the following PN metric in the unitary gauge
\be  %\nonumber
\begin{split}
\label{eq:PNunitg}
&g_{00}  =1+\Delta H+O(v^4),\\ 
&g_{ij}=\delta_{ij}\left(-1+\Delta H\right) -\frac{(\alpha+\beta+3\lambda)}{\beta+\lambda}\pd_j\pd_i H+O(v^4),\\
&g_{0i}=\frac{1}{4}(8+\alpha^{PPN}_1)(V^{PPN}_i+W^{PPN}_i)+O(v^4), \\
& \chi=O(v^4).
\end{split}
\ee

%%%%%%%%%%%%%%%%
\section{The Einstein-aether and the monopole}\label{app:monopole}
%%%%%%%%%%%%%%%%

In both khronometric and Einstein-aether theories, we compare the monopole contribution to the energy-loss formula
 in the limit for which the PPN parameters are identical to GR.  The free parameters of khronometric theory are $\a$, $\b$ and $\lambda$ and those of the Einstein-aether \cite{Jacobson:2008aj} are $c_i$ for $i=1,\dots,4$.   We have the correspondence\footnote{Recall that we are using the mostly minus signature. The mostly plus signature is used in \cite{Foster:2006az} and leads to a different correspondence between the parameters.} $c_1=0$, $c_2=\lambda$, $c_3=\beta$ and $c_4=\alpha$.  Notice that one less parameter is needed to define khronometric theory.  This is because the action of a hypersurface orthogonal aether (which is equivalent to khronometric theory \cite{Blas:2010hb}) contains a term that can be absorbed by the others, reducing the number of independent terms from four down to three.

Comparing the results of this paper and the work presented in \cite{Foster:2006az}, we see that the waveforms for the spin-0 and spin-2 modes are essentially identical.   The main difference comes from the expression for $Z$ of Eq. \eqref{eq:Z} (remind also footnote \ref{FosterM}).   Let $\tilde{Z}$ be the equivalent expression in \ae-theory,
\be
\tilde{Z}\equiv\frac{(c_{13}-1)(\tilde{\a}^{PPN}_1-2\tilde{\a}^{PPN}_2)}{3(c_{14}-2c_{13})},\label{Zae}
\ee
where $\tilde{\a}_1^{PPN}$, $\tilde{\a}_2^{PPN}$ are the Lorentz violating PPN parameters in \ae-theory and $c_{ij}=c_i+c_j$.  Then the khronometric expression for $Z$ is precisely  $\tilde{Z}$, but with $c_1=0$.   

The limit $\a_1^{PPN}=\a_2^{PPN}=0$ in khronometric theory can be achieved by setting $\alpha=2\beta$ and leads to $Z=1$.  By inspection of the energy-loss formula \eqref{eq:dotenerN}, we see that the monopole is proportional to $Z$ and therefore persists in this limit.    In generic \ae-theory, the equivalent limit that sets the PPN parameters to GR is given by
\be \label{eq:aetherlim}
c_2=\frac{-2c_1^2-c_1c_3+c_3^2}{3c_1}, \quad 
c_4=-\frac{c_3^2}{c_1}
\ee
and leads to $\tilde{Z}=0$.  The corresponding monopole is proportional to $\tilde{Z}$ and subsequently vanishes in this limit.  Therefore, the values of $Z$ and $\tilde{Z}$ explain the presence or absence of the monopole in the limit when the PPN parameters are identical to those of GR.

It is natural to ask if $\tilde{Z}$ can be tweaked so that \ae-theory has a monopole when $\tilde{\a}_1^{PPN}=\tilde{\a}_2^{PPN}=0$.  
A first possibility would be to consider the limit that ressembles khronometric theory, namely
 $c_1=0$ and $c_4=2c_3$. This leads to $\tilde{Z}=1$, like in khronometric theory, indicating that a monopole may be possible.  However, requiring only $c_1=0$ implies that $\a_1^{PPN}=8$.  One could try to set $c_1=0$ and  $c_3=0$ to get $\a_1^{PPN}=0$, but this case of \ae-theory has yet to be studied \cite{Jacobson:2008aj}. 
Alternatively, one may try to make the denominator in (\ref{Zae}) vanish to retrieve a finite limit. Setting  $c_{14}=2c_{13}$ yields $\tilde{Z}=1$.
However, the second condition in  (\ref{eq:aetherlim}) implies $c_1=c_3=c_4=0$, which is a singular limit for \ae-theory.

%%%%%%%%%%%%%%%%
\section{Notion of energy for an asymptotically flat spacetime}\label{ap:energy}
%%%%%%%%%%%%%%%%

To characterize the energy
carried away from a system by GWs, we
use a method different from the standard technique
defined in terms of the Landau-Lifshitz or related pseudotensors \cite{Will:Book,Maggiore:1900zz,Wein:Book,Szabados:2004vb}.
Here, instead of computing
the energy carried by GWs, we derive the loss of energy of the isolated system during the
process of gravitational radiation. This resembles
the definition of energy loss by the time variation of the Bondi-Sachs mass \cite{PoissonNotes,Szabados:2004vb}.
However, we will use a different notion of conserved energy that, to our knowledge,  was
first used in the context of GWs in  \cite{Foster:2006az}.  
This energy is well-defined for asymptotically flat spacetimes 
satisfying the boundary conditions (\ref{eq:fallPPN}), which we use to define  isolated sources.
  Its conservation follows from 
 the invariance of the asymptotic solution under time translations and it reduces
to the standard notion of energy for flat spacetime  \cite{Iyer:1994ys} (see also \cite{Regge:1974zd,Wald:1999wa}).
Since the method is not standard, this Appendix is devoted to presenting a succinct summary.
We encourage the reader to consult the original literature to complement it.

Given a Lagrangian density $L(\Phi)$ depending on some dynamical fields $\Phi$, we define its associated
4-form (we present the $3+1$ case) as
\be
\mathbf L(\Phi)=L(\Phi) \di^4 x . 
\nonumber
\ee
After integration by parts,  the first variation of the
previous form following from the variation $\delta \Phi$ can be expressed as, 
\be
\delta \mathbf L(\Phi)=\mathbf{E}_\Phi\delta \Phi+\di\mathbf \Theta_L(\Phi, \delta\Phi),\nonumber
\ee
where $\mathbf{E}_\Phi=0$ are the equations of motion of the theory. If
the  variation $\delta \Phi$ is a diffeomorphism generated by a vector field $\xi$, the previous variation should
correspond to the action of this transformation over $\mathbf L(\Phi)$, 
\be
\delta_\xi \mathbf L(\Phi)=\di (i_\xi \,\mathbf L),\nonumber
\ee
where $i_\xi \,\mathbf L$ refers to
 the contraction of the form $\mathbf L$ with the vector field $\xi$. 
 Define the Noether current 3-form associated to $\xi$ and $L(\Phi)$ as
\be
{\mathbf J}_L\equiv \mathbf \Theta_L(\Phi, \delta_\xi\Phi)-i_\xi \,\mathbf L.\label{eq:current}
\ee
This form is clearly closed when the equations of motion are satisfied. 
In practice, to find the components of the 3-form $\mathbf \Theta_L$,  notice that it is dual to a 1-form. In components
\be
 \mathbf \Theta_{L\m\n\r}=\mathbf{\epsilon}_{\a\m\n\r} \Theta_L^\a,\nonumber
\ee
where the index of $ \Theta_L^\a$ is risen with the metric $g^{\m\n}$ and
$\epsilon_{\a\m\n\r}$ are the components of the Levi-Civita $3$-form defined for the metric $g_{\m\n}$.
  From this definition it follows that
\be
\di \mathbf \Theta_L=\sqrt{-g}\nabla_\m  \Theta_L^\m\di^4 x=\pd_\m(\sqrt{-g} \Theta_L^\m)\di^4 x,\label{eq:conservcur}
\ee
from which one can easily identify the components of $\mathbf \Theta_L$.

To associate the flux generated  by $\xi$ to a Hamiltonian evolution from 
an initial hypersurface $\Sigma$, one must  
assume \cite{Iyer:1994ys,Wald:1999wa} that in the boundary of the initial hypersurface, denoted by $\pd \Sigma$, it is possible to find 
a 3-form $\mathbf B_L$
such that
\be
\delta \int_{\pd \Sigma} i_\xi \,\mathbf B_L = 
\int_{\pd \Sigma} i_\xi \, \mathbf \Theta_L.\nonumber
\ee
If such a current exists, the flux generated  by $\xi$ corresponds to the orbits generated by of the Hamiltonian
\be
\label{eq:hamil}
H_\xi \equiv \int_\Sigma  \mathbf J_L -
 \int_{\pd \Sigma} i_\xi  B_L.
\ee
Finally, since $\mathbf J_L$ is closed when the equations of motion are satisfied, it follows
that locally $\mathbf J_L =\di \mathbf Q_L$. Thus, when the 
equations of motion hold, $H_\xi$ can be written as a pure boundary term, 
\be
\label{eq:hamilf}
H_\xi =
 \int_{\pd \Sigma} \left(\mathbf Q_L- i_\xi \,B_L\right).
\ee

To define a canonical notion of energy, we shall now assume that  $\xi$ is an asymptotic time translation, with 
components $\xi^\m\to \delta^\m_0$ and that the asymptotic conditions on the dynamical fields have been specified in such a way that 
 the surface integrals appearing in Eq.~(\ref{eq:hamilf}) approach a finite limit. 
 The Hamiltonian then corresponds to the generator of time evolution.
 We define the canonical energy at a hypersurface slice of
  constant time $\Sigma_t$ to be \cite{Iyer:1994ys}
 \be
\label{eq:energy}
\mathcal  E _L=
 \int_{S^2_t}  \left(\mathbf Q_L- i_{\delta^\m_0} \,B_L\right),
\ee
where $S^2_t$ represents the boundary sphere at the boundary of  $\Sigma_t$.
Whenever $\mathcal  E _L$ is well-defined, it is a conserved quantity, 
and we can remove the $t$ label in $S_t^2$.

We now apply the previous formalism to our action (\ref{covar22}).
The hypersurface of constant time  corresponds to a sheet of the preferred foliation.
Even if not necessary, it is convenient to work with an action for which
\be
\label{eq:canbo}
\int_{S^2} i_\xi \, \mathbf \Theta_L=0.
\ee
 This equation is not satisfied  for the Einstein-Hilbert action part of (\ref{covar22})
 (see e.g. Eq.~(87) in \cite{Iyer:1994ys}).
As explained in \cite{Iyer:1994ys}, the existence of a background metric $\eta_{\m\n}$ makes it
possible to build a covariant action (which is required to get a conserved current (\ref{eq:current})) equivalent to Einstein-Hilbert and satisfying (\ref{eq:canbo}). Indeed, 
let us write $g_{\m\n}=\eta_{\m\n}+h_{\m\n}$ and consider $h_{\m\n}$ and $\eta_{\m\n}$ as independent
 dynamical fields. We can then add a boundary term invariant under diffeomorphisms to the action (\ref{covar22}) 
to yield
\be
\label{eq:actionp}
S'\equiv S+\frac{M_0^2}{2}\int \di^4 x\left(\sqrt{-g}\left((\Gamma^\a_{\m\n}-\bar\Gamma^\a_{\m\n})g^{\m\n}-(\Gamma^\m_{\m\n}-\bar\Gamma^\m_{\m\n}
)g^{\n\a}\right)\right),_{\a}\equiv
\int \di^4 x \,L',
\ee
where $\bar\Gamma^\m_{\s\n}$ refers to the connection compatible with the background metric $\eta_{\m\n}$.
The part corresponding to GR reads
\be
\label{eq:actprime}
S'_{\Gamma\Gamma}=-\frac{M_0^2}{2}\int \di^4 x\left[\sqrt{-g}g^{\m\r}\left(\Gamma^\a_{\r\n}\Gamma^\n_{\a\m}-\Gamma^\n_{\a\n}\Gamma^\a_{\m\r}\right)
+\left(\sqrt{-g}\left(\bar\Gamma^\a_{\m\n}g^{\m\n}-\bar\Gamma^\m_{\m\n}
g^{\n\a}\right)\right), _{\a}\right].
\ee
The equations of motion derived from varying the previous action with respect to $h_{\m\n}$ and $\eta_{\m\n}$ are the
same, as these fields appear only in the combination $g_{\m\n}$, except in the boundary term. 
 As a consequence, $\eta_{\m\n}$ can be considered to be Minkowski, and we can assume that the equations of motion
fix $h_{\m\n}$.

For the computation of $\mathbf J_{\Gamma\Gamma}$ 
corresponding to the action (\ref{eq:actprime}) and the vector field $\pd_t$, with components
$\delta^\m_0$, we first notice that $\pd_t$ is a Killing vector of $\eta_{\m\n}$,
\be
\delta_{\pd_t} \eta_{\m\n}=2\bar\nabla_{(\m}\eta_{\n)\a}\delta_0^\a=0,\nonumber
\ee
and the boundary term in Eq.~(\ref{eq:actprime}) does not contribute to $\mathbf J_{\Gamma\Gamma}$.  For the first term one finds
the corresponding current
\be
\label{eq:GRcurrent}
\begin{split}
 \Theta^\n_{\Gamma \Gamma}=
\frac{M_0^2}{4}\Big(\Gamma^\n_{\m\a}(g^{\m\a}g^{\r\s}\delta 
g_{\r\s}&-2g^{\m\r}g^{\a\tau}\delta g_{\tau \r}) +g^{\n\a}(2\Gamma_{\s\b}^\b g^{\r\s}\delta
g_{\r\a}-\Gamma_{\a\b}^\b g^{\r\s}\delta 
g_{\r\s} )\Big).
\end{split}
\ee
This  term is linear in the connection and does not depend on the derivative of $\delta g_{\m\n}$. To construct the conserved current, we use 
\be
\delta_{\pd_t} g_{\m\n}=2\nabla_{(\m}g_{\n)\a}\delta^\a_0=2g_{\a(\m}\Gamma_{\n)0}^\a,\nonumber
\ee
Thus, under the assumption that the fields fall-off at large distances 
as (\ref{eq:fallPPN}), the  current (\ref{eq:GRcurrent})  vanishes asymptotically
as $O(r^{-4})$, which means that its contribution to (\ref{eq:canbo}) cancels. Indeed the cancellation 
of the contribution to (\ref{eq:canbo}) holds in the more general situation where one considers
variations $\delta g_{\m\n}$ which do not change the asymptotic behaviour (\ref{eq:fallPPN}). 
 Finally,  the energy $\mathcal  E_{\Gamma\Gamma}$ derived from Eq.~(\ref{eq:actprime}) coincides with the ADM mass which also agrees
with the energy derived from the Landau-Lifshitz pseudotensor \cite{Iyer:1994ys,Sorkin:1991bw}.

The  term $S_\chi$ in the action (\ref{covar22}) yields a current
\be
\label{currentk}
\begin{split}
 \Theta_{\chi}^\n=-M_b^2\Big[&\left(\a a_\s \nabla_\m u^\s-\nabla_\r K^\r_{\phantom{\r}\m}\right)\frac{\mathcal P^{\n\m}}{\sqrt{X}}\delta \chi+K^{\n\r}\frac{\mathcal P_\r^\a}{\sqrt{X}}\pd_\a \delta \chi\\
&\quad \quad-\frac{1}{2}\left(\left[K^{\n\a}+K^{\a\n}\right]u^\s-K^{\a\s}u^\n-u^\a u^\s u_\r K^{\n\r}\right)\delta g_{\a\s}\Big].
\end{split}
\ee
Remember that
the  invariance  under diffeomorphisms is non-linearly realized\footnote{We could also work with the field $\varphi$ for which
$\delta_\xi \varphi=\xi^\m\pd_\m \varphi$.} on $\chi$
\be
\delta_\xi \chi=\xi^0+\xi^\m\pd_\m \chi.\nonumber
\ee
From Eq.~(\ref{eq:fallPPN}) this means
that $\delta_{\pd_t} \chi\sim O(1)$. Similarly $u_\a=\frac{\delta_{\a 0}}{\sqrt{g^{00}}}\sim \delta_{\a 0}+O(1/r)$. Thus, 
 $ \Theta^\n_{\chi}\sim O(r^{-3})$, which means that the contribution of this
term to (\ref{eq:canbo}) cancels.

 Finally, we find that the conserved energy (\ref{eq:hamil}) for the action (\ref{eq:actionp})
  inside a constant time hypersurface $\Sigma_t$ is given by
 \be
 \label{eq:energyf}
\mathcal  E =\int_{\Sigma_t} \di^3 x \sqrt{-g}\,\mathcal J^0_{S'}, 
 \ee
with $\mathcal J^0_{S'}$ representing the coordinates of the 1-form dual to the corresponding 3-form, Eq.~(\ref{eq:current}),
\be
\label{eq:conserf}
\begin{split}
\mathcal J^\n_{S'}&\equiv(\Theta_{\Gamma\Gamma}^\n+\Theta_{\chi}^\n)-\delta^\n_0 L'.
\end{split}
\ee
The contribution from the khronon action is simplified once 
one considers the equation of motion for $\chi$.  
 Indeed, $\Theta_\chi^\n$  in Eq.~(\ref{currentk}) includes a term 
\be
\left(\a a_\s \nabla_\m u^\s-\nabla_\r K^\r_{\phantom{\r}\m}\right)\frac{\mathcal P^{\n\m}}{\sqrt{X}}=J^\n,
\ee
where $J^\n$ is defined in (\ref{eq:eomk}). In the unitary gauge, this current is purely spatial, which means that this term does not contribute to 
(\ref{eq:energyf}).

We are eventually interested in the flux of energy loss through GWs, so we want to compute the quantity,
\be
\label{eq:energyfl}
\dot {\mathcal  E} =\int_\Sigma \di^3 x\, \sqrt{-g}\,\dot {\mathcal J}_{S'}^0=-\oint_{S_{\infty}^2}  \di \Omega \, \sqrt{-g}\, r^2\hat r^i \mathcal J_{S'}^i,
\ee
where we have used the fact that the current ${\mathcal J}_{S'}^\m$ is conserved on-shell, 
which is a consequence of $\mathbf J$ being closed and (\ref{eq:conservcur}). 
 The final ingredient is to evaluate $\mathcal J_{S'}^i$.  From Eq.~(\ref{eq:GRcurrent}), 
 \be
\begin{split}
\Theta^i_{\Gamma\Gamma}=
&\frac{M_b^2}{4}\dot h_{\a\b}\left[\eta^{\a\b}(\pd^\r h_\r^i-\pd^i h_{\s}^\s)-2\pd^\a h^{\b i}
+\pd^i h^{\a\b}+\eta^{\b i}\pd^\a h_{\s}^\s\right]+O(h^3).
\end{split}
\ee
For the khronon terms, we find that at quadratic order in the unitary gauge
\be
\begin{split}
\Theta^i_{\chi}=M_b^2\left[\bar K^{(\a i)}(\bar\Gamma^0_{\alpha 0}+\eta_{\a\r}
\bar \Gamma^\r_{00})-\bar K^{i0}\bar \Gamma^0_{00}\right].
\end{split}
\ee
From this expression it is clear that the notion of energy (\ref{eq:energyf}) is not well defined for spacetimes with radiation at
infinity satisfying conditions (\ref{eq:boundw}). This is an unphysical divergence, 
which is regularized for a flux of energy of finite duration \cite{Regge:1974zd}. 
For our purposes, it is enough to notice that the time variation (\ref{eq:energyfl})  (and hence the flux) is well defined 
for these boundary conditions. Also, only the part of the integral quadratic in perturbations 
does not vanish, which implies that the previous expressions are enough to compute the flux
of energy at infinity.
The steps to go from the previous formula to the final result (\ref{eq:dotener}) are 
explained in Sec. \ref{sec:energy}.

%%%%%%%%%%%%%%%%%%%%%%%%%%%


\begin{thebibliography}{99}
%%%%%%%%%%%%%%%%%%%%%%%%%%%


%\cite{Will:2005va}
\bibitem{Will:2005va}
  C.~M.~Will,
  %``The confrontation between general relativity and experiment,''
  Living Rev.\ Rel.\  {\bf 9} (2005) 3
  [arXiv:gr-qc/0510072].
  %%CITATION = 00222,9,3;%%
  
%\cite{ArkaniHamed:1998rs}
\bibitem{ArkaniHamed:1998rs}
  N.~Arkani-Hamed, S.~Dimopoulos and G.~R.~Dvali,
  %``The Hierarchy problem and new dimensions at a millimeter,''
  Phys.\ Lett.\  B {\bf 429} (1998) 263
  [arXiv:hep-ph/9803315].
  %%CITATION = PHLTA,B429,263;%%


%\cite{Dvali:2007hz}
\bibitem{Dvali:2007hz}
  G.~Dvali,
  %``Black Holes and Large N Species Solution to the Hierarchy Problem,''
  Fortsch.\ Phys.\  {\bf 58} (2010) 528
  [arXiv:0706.2050 [hep-th]].
  %%CITATION = FPYKA,58,528;%%


  
%\cite{Horava:2009uw}
\bibitem{Horava:2009uw}
  P.~Horava,
  %``Quantum Gravity at a Lifshitz Point,''
  Phys.\ Rev.\  {\bf D79 } (2009)  084008.
  [arXiv:0901.3775 [hep-th]].
  
  
%\cite{Blas:2010hb}
\bibitem{Blas:2010hb}
  D.~Blas, O.~Pujolas and S.~Sibiryakov,
  %``Models of non-relativistic quantum gravity: The Good, the bad and the
  %healthy,''
  JHEP {\bf 1104} (2011) 018
  [arXiv:1007.3503 []].
  %%CITATION = JHEPA,1104,018;%%



%\cite{Horava:2010zj}
\bibitem{Horava:2010zj}
  P.~Horava and C.~M.~Melby-Thompson,
  %``General Covariance in Quantum Gravity at a Lifshitz Point,''
  Phys.\ Rev.\  D {\bf 82} (2010) 064027
  [arXiv:1007.2410 [hep-th]].
  %%CITATION = PHRVA,D82,064027;%%


%\cite{Blas:2009qj}
\bibitem{Blas:2009qj}
  D.~Blas, O.~Pujolas and S.~Sibiryakov,
  %``Consistent Extension Of Horava Gravity,''
  Phys.\ Rev.\ Lett.\  {\bf 104} (2010) 181302
  [arXiv:0909.3525 [hep-th]].
  %%CITATION = PRLTA,104,181302;%%

 %\cite{Jacobson:2010mx}
\bibitem{Jacobson:2010mx}
  T.~Jacobson,
  %``Extended Horava gravity and Einstein-aether theory,''
  Phys.\ Rev.\  D {\bf 81}, 101502 (2010)
  [Erratum-ibid.\  D {\bf 82}, 129901 (2010)]
  [arXiv:1001.4823 [hep-th]].
  %%CITATION = PHRVA,D81,101502;%%  
  

%\cite{Blas:2009ck}
\bibitem{Blas:2009ck}
  D.~Blas, O.~Pujolas and S.~Sibiryakov,
  %``Comment on `Strong coupling in extended Horava-Lifshitz gravity',''
  Phys.\ Lett.\  B {\bf 688} (2010) 350
  [arXiv:0912.0550 [hep-th]].
  %%CITATION = PHLTA,B688,350;%%

%\cite{ArmendarizPicon:2010rs}
\bibitem{ArmendarizPicon:2010rs}
  C.~Armendariz-Picon, N.~F.~Sierra and J.~Garriga,
  %``Primordial Perturbations in Einstein-Aether and BPSH Theories,''
  JCAP {\bf 1007}, 010 (2010)
  [arXiv:1003.1283 [astro-ph.CO]].
  %%CITATION = JCAPA,1007,010;%%
  
%\cite{Kobayashi:2010eh}
\bibitem{Kobayashi:2010eh}
  T.~Kobayashi, Y.~Urakawa and M.~Yamaguchi,
  %``Cosmological perturbations in a healthy extension of Horava gravity,''
  JCAP {\bf 1004} (2010) 025
  [arXiv:1002.3101 [hep-th]].
  %%CITATION = JCAPA,1004,025;%%


%\cite{Will:Book}
\bibitem{Will:Book}
  C.~M.~Will,
  ``Theory and experiment in gravitational physics,''
%\href{http://www.slac.stanford.edu/spires/find/hep/www?irn=2796651}{SPIRES entry}
{\it  Cambridge, UK: Univ. Pr. (1993), 380 p.}% (ISBN-13: 978-0521439732).}


%\cite{Jacobson:2008aj}
\bibitem{Jacobson:2008aj}
  T.~Jacobson,
  %``Einstein-aether gravity: a status report,''
  PoS {\bf QG-PH} (2007) 020
  [arXiv:0801.1547 [gr-qc]].
  %%CITATION = POSCI,QG-PH,020;%%
  


%\cite{Blanchet:2006zz}
\bibitem{Blanchet:2006zz}
L.~Blanchet,
%``Gravitational Radiation from Post-Newtonian Sources and Inspiralling   Compact Binaries,''
Living Rev.\ Rel.\  {\bf 9} (2006) 4.
%%CITATION = 00222,9,4;%%

  
%\cite{Maggiore:1900zz}
\bibitem{Maggiore:1900zz}
  M.~Maggiore,
  ``Gravitational Waves. Vol. 1: Theory and Experiments,''
%\href{http://www.slac.stanford.edu/spires/find/hep/www?irn=7540906}{SPIRES entry}
{\it  Oxford University Press (2007), 572p. }%(ISBN-13: 978-0198570745)}.


%\cite{Taylor:1982zz}
\bibitem{Taylor:1982zz}
  J.~H.~Taylor and J.~M.~Weisberg,
  %``A new test of general relativity: Gravitational radiation and the binary
  %pulsar PS R 1913+16,''
  Astrophys.\ J.\  {\bf 253} (1982) 908.
  %%CITATION = ASJOA,253,908;%%


%\cite{Weisberg:2004hi}
\bibitem{Weisberg:2004hi}
  J.~M.~Weisberg,  {Nice}, D.~J. and J.~H.~Taylor,
  %``Timing Measurements of the Relativistic Binary Pulsar PSR B1913+16"
   Astrophys.\ J.\  {\bf 722} (2010) 1030,
  arXiv:1011.0718 [astro-ph.GA].

  


  
%\cite{Damour:1992we}
\bibitem{Damour:1992we}
  T.~Damour and G.~Esposito-Farese,
  %``Tensor multiscalar theories of gravitation,''
  Class.\ Quant.\ Grav.\  {\bf 9} (1992) 2093.
  %%CITATION = CQGRD,9,2093;%%

  
 %\cite{Cannella:2009he}
\bibitem{Cannella:2009he}
U.~Cannella, S.~Foffa, M.~Maggiore, H.~Sanctuary and R.~Sturani,
%``Extracting the Three- and Four-Graviton Vertices from Binary Pulsars and Coalescing Binaries,''
Phys.\ Rev.\ D {\bf 80} (2009) 124035
[arXiv:0907.2186 [gr-qc]].
%%CITATION = PHRVA,D80,124035;%%
 

%\cite{Foster:2006az}
\bibitem{Foster:2006az}
  B.~Z.~Foster,
  %``Radiation damping in Einstein-aether theory,''
  Phys.\ Rev.\  {\bf D73}, 104012 (2006).
  [gr-qc/0602004].
  
 
%\cite{Foster:2007gr}
\bibitem{Foster:2007gr}
  B.~Z.~Foster,
  %``strong-field effects on binary systems in Einstein-aether theory,''
  Phys.\ Rev.\  D {\bf 76} (2007) 084033
  [arXiv:0706.0704 [gr-qc]].
  %%CITATION = PHRVA,D76,084033;%%
  
%\cite{Barausse:2011pu}
\bibitem{Barausse:2011pu}
  E.~Barausse, T.~Jacobson, T.~P.~Sotiriou,
  %``Black holes in Einstein-aether and Horava-Lifshitz gravity,''
  Phys.\ Rev.\  {\bf D83}, 124043 (2011).
  [arXiv:1104.2889 [gr-qc]].


%\cite{Barausse:2011pu}
\bibitem{BlasSibi}
D.~Blas and S.~Sibiryakov, 
{\it in preparation.}


%\cite{Eling:2007xh}
\bibitem{Eling:2007xh}
  C.~Eling, T.~Jacobson and M.~Coleman Miller,
  %``Neutron stars in Einstein-aether theory,''
  Phys.\ Rev.\  D {\bf 76} (2007) 042003
  [Erratum-ibid.\  D {\bf 80} (2009) 129906]
  [arXiv:0705.1565 [gr-qc]].
  %%CITATION = PHRVA,D76,042003;%%
  
 

%\cite{Kimpton:2010xi}
\bibitem{Kimpton:2010xi}
I.~Kimpton and A.~Padilla,
%``Lessons from the Decoupling Limit of Horava Gravity,''
JHEP {\bf 1007} (2010) 014
[arXiv:1003.5666 [hep-th]].
%%CITATION = JHEPA,1007,014;%%

 
 
%\cite{Blas:2009yd}
\bibitem{Blas:2009yd}
D.~Blas, O.~Pujolas and S.~Sibiryakov,
%``On the Extra Mode and Inconsistency of Horava Gravity,''
JHEP {\bf 0910} (2009) 029
[arXiv:0906.3046 [hep-th]].
%%CITATION = JHEPA,0910,029;%%

%\cite{Blanchet:1989ki}
\bibitem{Blanchet:1989ki}
  L.~Blanchet and T.~Damour,
  %``POSTNEWTONIAN GENERATION OF GRAVITATIONAL WAVES,''
  Annales Poincare Phys.\ Theor.\  {\bf 50} (1989) 377.
  %%CITATION = AHPAA,50,377;%%

%\cite{PoissonNotes}
\bibitem{PoissonNotes}
E.~Poisson,
``Post-Newtonian theory for the common reader,''
Lecture notes (July 2007).

%\cite{Pati:2002ux}
\bibitem{Pati:2002ux}
M.~E.~Pati and C.~M.~Will,
%``Post-Newtonian Gravitational Radiation and Equations of Motion via  Direct   Integration of the Relaxed Einstein Equations. Ii: Two-Body  Equations of   Motion to Second Post-Newtonian Order, and  Radiation-Reaction to 3.5   Post-Newton,''
Phys.\ Rev.\  D {\bf 65} (2002) 104008
[arXiv:gr-qc/0201001].
%%CITATION = PHRVA,D65,104008;%%

  
%\cite{Jacobson:2004ts}
\bibitem{Jacobson:2004ts}
  T.~Jacobson and D.~Mattingly,
  %``Einstein-Aether waves,''
  Phys.\ Rev.\  D {\bf 70} (2004) 024003
  [arXiv:gr-qc/0402005].
  %%CITATION = PHRVA,D70,024003;%%

%\cite{Wein:Book}
\bibitem{Wein:Book}
  S.~Weinberg, 
   ``Gravitation and Cosmology: Principles and Applications of the General Theory of Relativity,''
{\it John Wiley \& Sons, Inc. (1972), 657 p. }%(ISBN-13: 978-0471925675)}.
  

%%\cite{Blanchet:2002av}
%\bibitem{Blanchet:2002av}
%  L.~Blanchet,
%  %``Gravitational radiation from post-Newtonian sources and inspiralling
%  %compact binaries,''
%  Living Rev.\ Rel.\  {\bf 5} (2002) 3
%  [arXiv:gr-qc/0202016].
%  %%CITATION = 00222,5,3;%%
  
 %\cite{Szabados:2004vb}
\bibitem{Szabados:2004vb}
  L.~B.~Szabados,
  %``Quasi-Local Energy-Momentum and Angular Momentum in GR: A Review Article,''
  Living Rev.\ Rel.\  {\bf 7} (2004) 4.
  %%CITATION = 00222,7,4;%%
 
%\cite{Foster:2005fr}
\bibitem{Foster:2005fr}
B.~Z.~Foster,
%``Noether Charges and Black Hole Mechanics in Einstein-Aether Theory,''
Phys.\ Rev.\  D {\bf 73} (2006) 024005
[arXiv:gr-qc/0509121].
%%CITATION = PHRVA,D73,024005;%%
 
 
%\cite{Iyer:1994ys}
\bibitem{Iyer:1994ys}
  V.~Iyer and R.~M.~Wald,
  %``Some properties of Noether charge and a proposal for dynamical black hole
  %entropy,''
  Phys.\ Rev.\  D {\bf 50} (1994) 846
  [arXiv:gr-qc/9403028].
  %%CITATION = PHRVA,D50,846;%%
  
%\cite{Sorkin:1991bw}
\bibitem{Sorkin:1991bw}
  R.~D.~Sorkin,
  %``The Gravitational Electromagnetic Noether Operator And The Second Order
  %Energy Flux,''
  Proc.\ Roy.\ Soc.\ Lond.\  A {\bf 435} (1991) 635.
  %%CITATION = PRSLA,A435,635;%%

%\cite{Regge:1974zd}
\bibitem{Regge:1974zd}
  T.~Regge and C.~Teitelboim,
  %``Role Of Surface Integrals In The Hamiltonian Formulation Of General
  %Relativity,''
  Annals Phys.\  {\bf 88} (1974) 286.
  %%CITATION = APNYA,88,286;%%

%\cite{Peters:1963ux}
\bibitem{Peters:1963ux}
  P.~C.~Peters and J.~Mathews,
  %``Gravitational radiation from point masses in a Keplerian orbit,''
  Phys.\ Rev.\  {\bf 131} (1963) 435.
  %%CITATION = PHRVA,131,435;%%


 %\cite{Weisberg:1981bh}
\bibitem{Weisberg:1981bh}
  J.~M.~Weisberg and J.~H.~Taylor,
  %``Gravitational Radiation From An Orbiting Pulsar,''
  Gen.\ Rel.\ Grav.\  {\bf 13} (1981) 1.
  %%CITATION = GRGVA,13,1;%%

%\cite{Eardley75}
\bibitem{Eardley75}
  D.~M.~Eardley, 
  %``Observable effects of a scalar gravitational field in a binary pulsar,''
  Ap. J. {\bf 196} L59-L62 (1975).


%\cite{Damour:1996ke}
\bibitem{Damour:1996ke}
  T.~Damour and G.~Esposito-Farese,
  %``Tensor-scalar gravity and binary-pulsar experiments,''
  Phys.\ Rev.\  D {\bf 54} (1996) 1474
  [arXiv:gr-qc/9602056].
  %%CITATION = PHRVA,D54,1474;%%


%\cite{EspositoFarese:2004cc}
\bibitem{EspositoFarese:2004cc}
G.~Esposito-Farese,
%``Tests of Scalar-Tensor Gravity,''
AIP Conf.\ Proc.\  {\bf 736} (2004) 35
[arXiv:gr-qc/0409081].
%%CITATION = APCPC,736,35;%%




%\cite{Goldberger:2004jt}
\bibitem{Goldberger:2004jt}
W.~D.~Goldberger and I.~Z.~Rothstein,
%``An Effective Field Theory of Gravity for Extended Objects,''
Phys.\ Rev.\  D {\bf 73} (2006) 104029
[arXiv:hep-th/0409156].
%%CITATION = PHRVA,D73,104029;%%


%\cite{Porto:2006bt}
\bibitem{Porto:2006bt}
R.~A.~Porto and I.~Z.~Rothstein,
%``The Hyperfine Einstein-Infeld-Hoffmann Potential,''
Phys.\ Rev.\ Lett.\  {\bf 97} (2006) 021101
[arXiv:gr-qc/0604099].
%%CITATION = PRLTA,97,021101;%%


%\cite{Porto:2005ac}
\bibitem{Porto:2005ac}
R.~A.~Porto,
%``Post-Newtonian Corrections to the Motion of Spinning Bodies in Nrgr,''
Phys.\ Rev.\ D {\bf 73} (2006) 104031
[arXiv:gr-qc/0511061].
%%CITATION = PHRVA,D73,104031;%%


%\cite{Goldberger:2009qd}
\bibitem{Goldberger:2009qd}
W.~D.~Goldberger and A.~Ross,
%``Gravitational Radiative Corrections from Effective Field Theory,''
Phys.\ Rev.\ D {\bf 81} (2010) 124015
[arXiv:0912.4254 [gr-qc]].
%%CITATION = PHRVA,D81,124015;%%




%\cite{Carroll:2004ai}
\bibitem{Carroll:2004ai}
S.~M.~Carroll and E.~A.~Lim,
%``Lorentz-Violating Vector Fields Slow the Universe Down,''
Phys.\ Rev.\  D {\bf 70} (2004) 123525
[arXiv:hep-th/0407149].
%%CITATION = PHRVA,D70,123525;%%



%\cite{Blas:2011en}
\bibitem{Blas:2011en}
  D.~Blas, S.~Sibiryakov,
  %``Technically natural dark energy from Lorentz breaking,''
  JCAP {\bf 1107}, 026 (2011).
  [arXiv:1104.3579 [hep-th]].

%\cite{Yunes:2009ke}
\bibitem{Yunes:2009ke}
  N.~Yunes and F.~Pretorius,
  %``Fundamental Theoretical Bias in Gravitational Wave Astrophysics and the
  %Parameterized Post-Einsteinian Framework,''
  Phys.\ Rev.\  D {\bf 80} (2009) 122003
  [arXiv:0909.3328 [gr-qc]].
  %%CITATION = PHRVA,D80,122003;%%

%\cite{Yunes:2010qb}
\bibitem{Yunes:2010qb}
  N.~Yunes and S.~A.~Hughes,
  %``Binary Pulsar Constraints on the Parameterized post-Einsteinian
  %Framework,''
  Phys.\ Rev.\  D {\bf 82} (2010) 082002
  [arXiv:1007.1995 [gr-qc]].
  %%CITATION = PHRVA,D82,082002;%%
 
%\cite{DelPozzo:2011pg}
\bibitem{DelPozzo:2011pg}
  W.~Del Pozzo, J.~Veitch, A.~Vecchio,
  %``Testing General Relativity using Bayesian model selection: Applications to observations of gravitational waves from compact binary systems,''
  Phys.\ Rev.\  {\bf D83 } (2011)  082002.
  [arXiv:1101.1391 [gr-qc]].




%\cite{Foster:2005dk}
\bibitem{Foster:2005dk}
B.~Z.~Foster and T.~Jacobson,
%``Post-Newtonian Parameters and Constraints on Einstein-Aether Theory,''
Phys.\ Rev.\  D {\bf 73} (2006) 064015
[arXiv:gr-qc/0509083].
%%CITATION = PHRVA,D73,064015;%%



  


%\cite{Wald:1999wa}
\bibitem{Wald:1999wa}
  R.~M.~Wald and A.~Zoupas,
  %``A General Definition of "Conserved Quantities" in General Relativity and
  %Other Theories of Gravity,''
  Phys.\ Rev.\  D {\bf 61} (2000) 084027
  [arXiv:gr-qc/9911095].
  %%CITATION = PHRVA,D61,084027;%%


\end{thebibliography}
\end{document}